\documentclass[runningheads,orivec]{llncs}
\usepackage[T1]{fontenc}
\usepackage{graphicx}
\usepackage{cite}
    
\usepackage{anyfontsize}

\usepackage{stmaryrd}
\SetSymbolFont{stmry}{bold}{U}{stmry}{m}{n}
\usepackage{xcolor}
\usepackage{listings}
\usepackage{verbatim}
\lstdefinelanguage{scala}{
  alsoletter={@},
  morekeywords={abstract, case, catch, choose, class, def, do, else, extends, final, finally, for, if, implicit, import, match, new, null, object, let, in, be, lazy,holds,@erasable, override, package, private, protected, requires, ensures, decreases, return, sealed, super, then, this, throw, trait, try, type, val, var, while, yield, assert, _, return, require, ensuring, enum, end, @ghost, using},
  sensitive=true,
  morecomment=[l]{//},
  morecomment=[s]{/*}{*/},
  morestring=[b]",
  literate={{=>}{$\Rightarrow$\,}2%
            {->}{$\rightarrow$\,}2%
            {<=}{$\le$\,}2%
            {>=}{$\ge$\,}2%
            {<-}{$\leftarrow$\,}2%
            {++}{$\concat$\,}2%
            }
}

\newcommand{\codestyle}{\small\sffamily}

\newcommand{\commentstyle}{\small\color{teal}\textsl}

\newcommand{\asncommentstyle}{\small\color{blue}}

\lstdefinelanguage{asn}{
  keywords={
  OF, SEQUENCE, SIZE, CHOICE, INTEGER
  },
  sensitive=true,
  morecomment=[l]{//},
  morecomment=[s]{/*}{*/},
  morecomment=[s]{[}{]},
  morestring=[b]",
  columns=flexible,
  commentstyle=\asncommentstyle,
  literate={{<}{$\langle$}1%
            {>}{$\rangle$}1%
            {-}{$\mhyphen$}1%
            }
}

\usepackage{amsmath}
\usepackage{amsfonts}
\usepackage{graphicx}
\usepackage{epstopdf}
\usepackage{xspace}
\usepackage{hyperref}

\usepackage{bbding}

\usepackage{color}

\begin{document}
\sloppy
\title{Formally Verified Linear-Time Invertible Lexing}
\author{Samuel Chassot\orcidID{0009-0000-9751-9252} \Envelope \and Viktor Kun\v{c}ak\orcidID{0000-0001-7044-9522}}
%\author{Submission \#105. Reference for Artifact Evaluation}

%
% \authorrunning{S. Chassot et al.}
%\authorrunning{Submission \#105}
\institute{EPFL, Switzerland \\
\email{\{samuel.chassot \Envelope,viktor.kuncak\}@epfl.ch}}
\maketitle              % typeset the header of the contribution
\newcommand{\smartparagraph}[1]{\smallskip\noindent\textbf{#1.\,}}
\newcommand{\List}{\textit{List}}
\newcommand{\Set}{\textit{Set}}
\newcommand{\Context}{\textit{Context}}
\newcommand{\match}{\textit{match}}
\newcommand{\matchR}{\textit{matchR}}
\newcommand{\matchRSpec}{\textit{matchRSpec}}
\newcommand{\focus}{\textit{focus}}
\newcommand{\unfocus}{\textit{unfocus}}
\newcommand{\matchZ}{\textit{matchZ}}
\newcommand{\depth}{\textit{depth}}
\newcommand{\Union}{\textit{Union}}
\newcommand{\prefixMatch}{\textit{prefixMatch}}
\newcommand{\findLongestMatch}{\textit{findLongestMatch}}
\newcommand{\findLongestMatchTR}{\textit{findLongestMatchTR}}
\newcommand{\lex}{\textit{lex}}
\newcommand{\maxPrefixOneRule}{\textit{maxPrefixOneRule}}
\newcommand{\maxPrefix}{\textit{maxPrefix}}
\newcommand{\print}{\textit{print}}
\newcommand{\printWithSep}{\textit{printWithSep}}
\newcommand{\transform}{\textit{transform}}
\newcommand{\printWithSepWhenNeeded}{\textit{printWithSepWhenNeeded}}
\newcommand{\furthestNullablePosition}{\textit{furthestNullablePosition}}
\newcommand{\furthestNullablePos}{\textit{furthestNullablePos}}
\newcommand{\sep}{\textit{sep}}
\newcommand{\sepL}{\textit{sepL}}
\newcommand{\prefixSet}{\textit{prefixSet}}
\newcommand{\PrintableTokens}{\textit{PrintableTokens}}
\newcommand{\Character}{\textit{Character}}
\newcommand{\ElementMatch}{\textit{ElementMatch}}
\newcommand{\update}{\textit{update}}
\newcommand{\remove}{\textit{remove}}
\newcommand{\head}{\textit{head}}
\newcommand{\charsOf}{\textit{charsOf}}
\newcommand{\concat}{\mathop{\texttt{++}}}

\newcommand{\ZipLex}{\textsc{ZipLex}\xspace}

\newcommand{\vbatim}{\text{Verbatim{\tt ++}}\xspace}

\newcommand{\ffbox}[1]{%
  {% open a group for a local setting
   \setlength{\fboxsep}{-2\fboxrule}% the rule will be inside the box boundary
\fbox{\hspace{1.2pt}\strut{\,#1\,}\hspace{1.2pt}}% print the box, with some padding at the left and right
  }% close the group
}

\newcommand{\tkn}[1]{\ffbox{\texttt #1}}

% Safe fallback when not using acmart
\providecommand{\Description}[1]{}
\begin{abstract}
%We present a formally verified invertible lexer that correctly and efficiently implements regular expression token definitions and follows the maximal munch principle.

We present \ZipLex, a verified framework for invertible linear-time lexical analysis following the 
longest match (maximal munch) semantics.
Unlike past verified lexers that focus only on satisfying the semantics of regular expressions and the longest match property,
\ZipLex also guarantees that lexing and printing are mutual inverses. Thanks to verified memoization, it
also ensures that the lexical analysis of a string is
linear in the size of the string.
Our design and implementation rely on two sets of ideas: (1) a new abstraction of token sequences that captures the separability of tokens in a sequence while supporting their efficient manipulation, and
(2) a combination of verified data structures and optimizations, including Huet's zippers and memoization with a standalone verified imperative hash table. Our hash table offers competitive performance as shown by our evaluation.
We implemented and verified \ZipLex using the Stainless deductive verifier for Scala.
Our evaluation demonstrates that \ZipLex supports realistic applications such as
JSON processing and lexers of programming languages, and behaves linearly even in cases that make flex-style approaches quadratic. \ZipLex is two orders of magnitude faster than \vbatim, showing that verified invertibility and linear-time algorithms can be developed without prohibitive cost. Compared to Coqlex, \ZipLex also offers linear (instead of quadratic) time lexing, and is the first lexer that comes with invertibility proofs for printing token sequences.

\sloppy

\end{abstract}

% CAV26 REQUIREMENTS
% Regular Papers (18 pages max, must be anonymized)

% Intro needs to be exciting, contain examples, and get to the point as quickly as possible
\section{Introduction}

Lexical analysis, or \emph{lexing}, is the first step in most parsing pipelines, including compilers and data analysis tools. A lexical analyzer (\emph{lexer}) transforms a raw sequence of characters into a sequence of structured tokens that feeds into subsequent phases. Given their critical role, lexers should satisfy strong correctness guarantees. Yet lexers often remain trusted components: even in the CompCert verified C compiler~\cite{leroy:hal-01238879}, the lexer is part of the trusted computing base and is not itself verified~\cite{leroy_compcert_2024}. Verified lexer generators such as Coqlex~\cite{ouedraogo2023coqlex} and the pioneering Verbatim frameworks \cite{egolf_verbatim_2021,egolf2022verbatim++} demonstrate that formal verification of core lexing properties is feasible, but challenges remain in performance, as well as in supporting use cases that involve printing tokens back to text.

In many applications, lexers are used together with printing, for instance when manipulating syntax trees in IDE refactoring engines, program synthesis tools, or compiler pipelines. In such settings, it is essential that printing a sequence of tokens produces a string whose subsequent lexical analysis yields back the same tokens. Without this guarantee, information may be silently lost.

We refer to this property as \emph{invertibility}. Invertible lexing ensures that alternating lexing and printing preserves information exactly, and can be seen as a prerequisite for fully invertible parsing and pretty-printing frameworks. While researchers studied invertible parsing~\cite{Rendel_2010_parsing_pretty_printing}, invertibility has not been investigated for lexical analysis, where longest match makes common parsing techniques inadequate. Existing frameworks that support invertible parsing, such as Grammatical Frameworks~\cite{ranta-2011}, do not enforce invertibility of user-defined lexers.

\emph{We present} \ZipLex, \emph{a framework for verified and efficient invertible lexing.} \ZipLex provides a fully verified lexer and regular expression engine based on Brzozowski's derivatives \cite{brzozowski1964derivatives,owens2009regular}. To reason about safe printing of token sequences, we develop a simple and general abstraction for reasoning about R-paths (sequences constrained by a given binary relation). We use this abstraction to characterize sequences of tokens that can be printed and re-lexed without merging adjacent tokens. Our predicates are preserved under slicing and concatenation, enabling efficient manipulation of token sequences while maintaining invertibility.

To achieve acceptable performance, \ZipLex incorporates verified optimizations, including memoization following Reps~\cite{DBLP:journals/toplas/Reps98}, yielding linear-time lexing in the length of the input string, and the use of Huet's zippers~\cite{huet1997zipper} and regular expression derivatives for regular expression matching. Memoization makes use of a verified and reusable memoization framework including a standalone verified mutable hash table, building on top of previous work~\cite{chassot2024verifying}, which offers competitive performance as shown by our evaluation. We implemented all components in Scala and verified them using Stainless \cite{DBLP:journals/pacmpl/HamzaVK19}. The verified code remains compatible with the standard Scala compilation and build toolchain.

We evaluate \ZipLex on several scenarios, including a JSON lexer and a JSON object-sorting application built on top of it. Our goal is to support high-fidelity applications that require strong guarantees about correctness and reversibility, such as fully verified compilers and verified communication protocols. 
%description languages based on ASN.1~\cite{bucev2025formally}.

\smartparagraph{Illustration of the challenge}
Mentioned as early as~\cite{cattell1978formalization}, the maximal munch principle, or \emph{longest match semantics}, is the classical correctness criterion for lexers. It states that each produced token should be the longest possible at a given position, given the input string and the token definitions. While sufficient for correct tokenization in isolation, longest match alone is insufficient to guarantee correct interaction between lexing and printing.

Consider the Scala-like statement \lstinline{val x = 1}, whose token sequence is
% \[
\tkn{val}\ \tkn{\ }\ \tkn{x}\ \tkn{\ }\ \tkn{=}\ \tkn{\ }\ \tkn{1}.
% \]
Suppose now that a refactoring tool removes whitespace tokens around \tkn{=}, yielding
% \[
\tkn{val}\ \tkn{\ }\ \tkn{x}\ \tkn{=}\ \tkn{1}.
% \]
Printing this sequence produces \lstinline{val x=1}, which is lexed back into the same sequence of tokens, up to whitespace, so no essential information is lost.
In contrast, consider instead the statement \lstinline{val x_ = 1}. Removing whitespace around the equality produces
% \[
\tkn{val}\ \tkn{\ }\ \tkn{x\_}\ \tkn{=}\ \tkn{1},
% \]
which prints as \lstinline{val x_=1}. Lexing this string yields
% \[
\tkn{val}\ \tkn{\ }\ \tkn{x\_=}\ \tkn{1},
% \]
because $\tkn{x\_=}$ is a valid identifier (e.g., in Scala) and is recognized by longest match. The printed output therefore does not lex back to the original token sequence, breaking our expectations.
Maintaining all whitespace tokens would avoid this issue, but this is incompatible with pretty printers and refactoring tools whose purpose is precisely to modify whitespace according to formatting conventions. The same problem can arise when reordering sub-sequences, for example, moving function definitions around in a source file.

The central problem we address is therefore:
\emph{How can we support printing and lexing while formally guaranteeing that no information is lost?}
To formalize this problem, we define \emph{separable tokens} and \emph{invertibility}.

\smartparagraph{Invertibility as a design goal}
We say that a lexer is \emph{invertible} if both:
\[
\forall\, ts.\ \lex(\print(ts)) = ts \quad\text{and}\quad \forall s.\ \print(\lex(s)) = s
\]
Note that if $\forall s.\ \print(\lex(s)) = s$ then
$\lex(\print(ts)) = ts$ whenever $\exists s.\, ts = lex(s)$.
We define a \emph{separability condition}, $\sep$, as any predicate on token sequences $ts$ such that 
$
\forall\, ts.\ \sep(ts) \implies \lex(\print(ts)) = ts.
$
Our goal is to design separability conditions fulfilling two competing requirements: (1) they are flexible enough for practical use and (2) they can be checked and preserved efficiently when token sequences are transformed in user code.

\smartparagraph{Contributions}
Our contributions include:
\begin{enumerate}
    \item a definition of separability $\sep(ts)$ and efficient mechanisms to enforce it;
    \item the design of a lexing framework that implements: (i) invertible printing (ii) regular-expression-based token definitions (iii) longest match semantics, and (iv) linear-time lexing thanks to memoization;
    \item a library of efficient verified data structures including a tree-based structure for sequences using immutable arrays with a zero-copy slicing operation in the leaves, a mutable hash table for arbitrary keys based on verified LongMap~\cite{chassot2024verifying}, and a reusable verified memoization framework.
    \item the concrete realization and deductive verification of a verified lexer, \ZipLex, and its performance evaluation showing its efficiency, available as a Docker image~\cite{chassot_formally_2026}, and in Bolts repository~\cite{epfl_larabolts_2026} under \verb|lexers/regex|.
    \end{enumerate}
%Our paper presents a design and deductive verification of a lexer meeting these objectives while being reasonably efficient in practice.

\subsection{Related Work}

We are aware of no existing work on lexers with \emph{verified invertibility}; we survey works that verify weaker or related properties.

\smartparagraph{Verified lexing}
Several verified lexer generators have been proposed. Coqlex~\cite{ouedraogo2023coqlex} provides a verified lexer generator in Rocq with a user experience close to OCamllex. It relies on Brzozowski's derivatives and supports additional constructs such as negation and difference, but is restricted to ASCII strings and depends on an unverified code-generation step. Verbatim~\cite{egolf_verbatim_2021}, later extended as \vbatim~\cite{egolf2022verbatim++}, also uses derivatives, with \vbatim introducing a translation to DFAs and memoization to improve performance, yielding a worst-case complexity of $O(n\log n)$. Our experiments in Section~\ref{sec:performance} show the performance advantages of our lexer, \ZipLex.

Nipkow~\cite{nipkow1998verified} presents an early verified lexer in Isabelle/HOL supporting arbitrary alphabets, based on an NFA-style algorithm that is not directly executable. In contrast, our technique avoids automata construction altogether, supports arbitrary alphabets, and achieves linear-time lexing through a verified variant of memoization~\cite{DBLP:journals/toplas/Reps98}, without an upfront preprocessing overhead.

\smartparagraph{Regular expressions and grammars}
Brzozowski's derivatives \cite{brzozowski1964derivatives} have been extensively studied as a basis for regular expression matching. Owens et al.~\cite{owens2009regular} re-established derivatives as a practical technique, and later work demonstrated that derivative-based engines can achieve state-of-the-art performance~\cite{varatalu2025re}, albeit without formal verification.

Edelmann's thesis~\cite{edelmann_efficient_2021} explores derivatives for LL(1) parsing~\cite{edelmann_2020PLDI_zippy} and lexing, introducing an optimization based on Huet's zippers~\cite{huet1997zipper}. While that work provides high-level proofs in Rocq, our effort with \ZipLex bridges the gap to implementation, offering a verified executable zipper-based matching engine. The implementation of zipper-based regular expression representation and derivative computation in Edelmann's work~\cite{edelmann_2020PLDI_zippy} served as a starting point for \ZipLex. We did not make use of the associated Rocq proofs, though they provide useful insights. Derivatives and zippers have also been used to parse  grammars~\cite{DBLP:conf/oopsla/BrachthauserRO16,DBLP:journals/pacmpl/DarraghA20}, but without end-to-end formal verification.
Chattopadhyay et al.~\cite{chattopadhyay_verified_2025} verify regular expression matching with lookarounds in Rocq, and De Santo et al.~\cite{DBLP:journals/pacmpl/SantoBP24} formalize JavaScript regular expression semantics. These works target richer regex features than \ZipLex but do not aim for executable longest-match lexing.

\smartparagraph{Invertibility in parsing and serialization}
Invertible parsing and pretty printing have been studied extensively, notably by Rendel et al.~\cite{Rendel_2010_parsing_pretty_printing}, who enforce invertibility by restricting transformations to isomorphisms but without instantiating it to lexical analysis. 
%A simple aspect of \ZipLex applies a related idea by requiring injective token transformations to guarantee invertible printing.
%
A related problem is serialization and deserialization; Bucev et al.~\cite{bucev2025formally} present a proof-producing ASN.1/SCC compiler in Scala, where, given a data format description, serialization and deserialization functions are generated together with proofs of invertibility accepted by Stainless. Lossless image compression also has a natural specification using mutually inverse functions
%; an example of a verified implementation is one for \emph{Quite OK Image Format} 
\cite{DBLP:conf/fmcad/BucevK22}.

%Finally, Madhavan et al.~\cite{MadhavanKuncak17Memoization} study verification of memoization and resource bounds in higher-order programs, informing our approach to efficient verified execution.

\section{Usage Example and \ZipLex API}
\label{sec:usage-examples}

\begin{figure}[hbt]
    \centering
    \includegraphics[width=0.63\linewidth,alt={Diagram showing steps in execution of a JSON sorting application. Step 1 shows the text representation of a JSON array containing objects, step 2 shows the result of applying lexical analysis, step 3 shows the tokens wrapped in a box representing the R-Path predicate being valid, step 4 shows the objects after splitting in gray boxes, step 5 shows the JSON objects after sorting individually wrapped in gray boxes, step 6 shows these objects concatenated as a JSON array in a single gray box, and step 7 shows them as text.}]{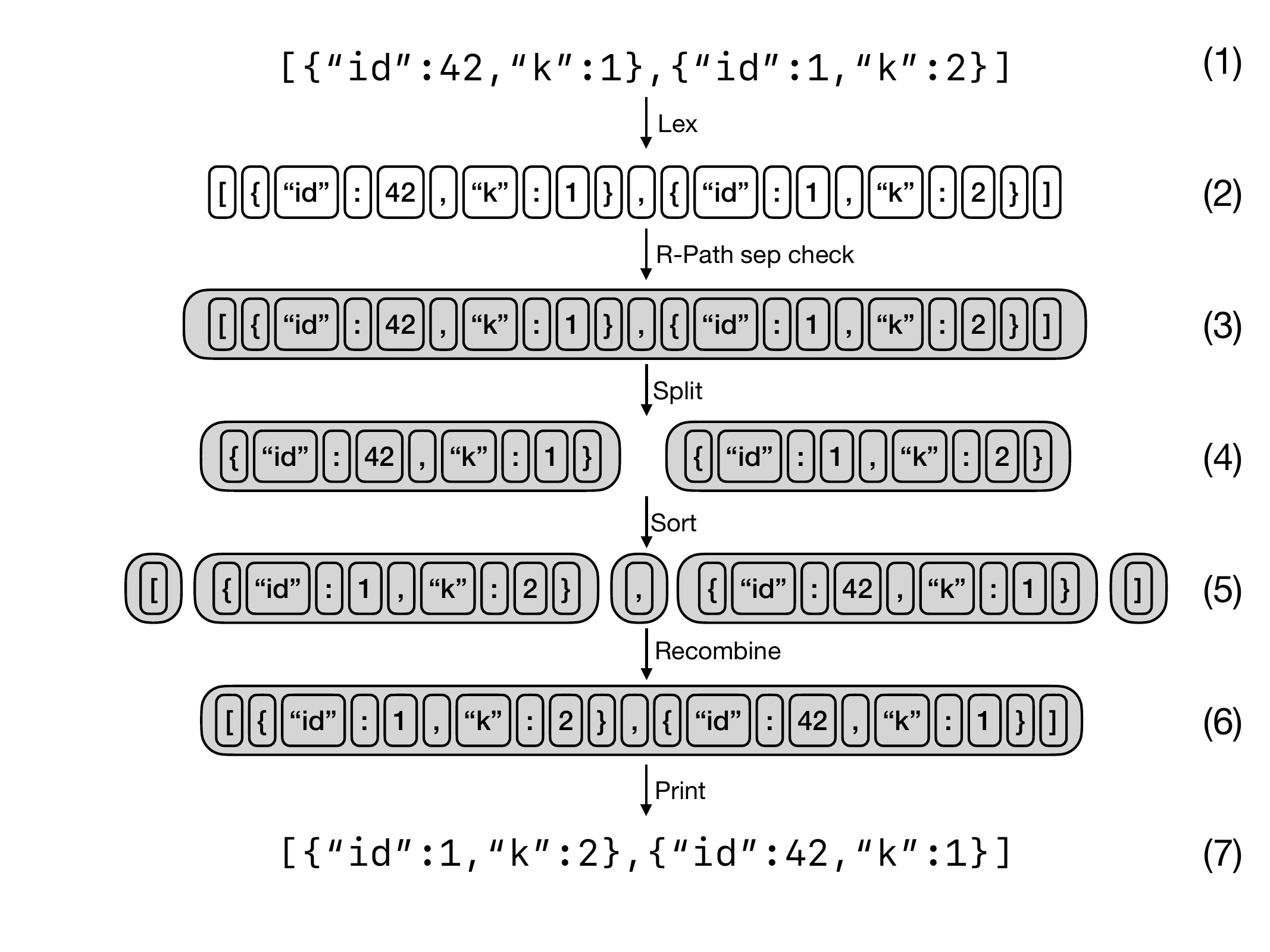}
    \caption{JSON sorting application. Outer gray boxes are \lstinline{PrintableTokens} instances.}
    \label{fig:json-sorting-algorithm-diagram}
\end{figure}

To demonstrate the verified abstraction, we present \ZipLex using a JSON lexer and a simple application that sorts the objects within a JSON array (Figure~\ref{fig:json-sorting-algorithm-diagram}). The implementation is included in the artifact, along with additional example lexers including one for a Python subset and one for Amy, a teaching language.

\smartparagraph{JSON application}
The application proceeds as follows. It lexes the input JSON file, and wraps the resulting token sequence in a \lstinline{PrintableTokens} instance, establishing the separability invariant. The application then shallowly parses the sequence to identify JSON objects and extracts their \texttt{"id"} fields, yielding object slices as \lstinline{PrintableTokens} instances. It then sorts these objects by the extracted identifiers and concatenates the resulting \lstinline{PrintableTokens} instances to form a single \lstinline{PrintableTokens} instance. Finally, the application prints back the object sequence to a character stream.

All token-sequence manipulations are performed through \lstinline{PrintableTokens}, which preserves separability under slicing and concatenation, requiring only a constant-time runtime check at concatenation boundaries. By verifying this program in Stainless, we statically guarantee that printing the final token sequence and lexing it again yields the same tokens.

\smartparagraph{Lexer interface and usage}
%\label{subsec:usage-examples-lexer-interface}
\ZipLex performs lexical analysis using data structures generated on-the-fly, without a code-generation phase used in, e.g., Ocamllex~\cite{Ocamllex}. Users define rules by associating regular expressions with token constructors, and invoke the main $\lex$ function with a rule list and an input sequence. The result is a sequence of tokens and a possibly untokenized suffix, corresponding to the uniform tokenization problem~\cite{Li_Mamouras_2025}. At each step, $\lex$ selects the longest prefix matched by any rule, breaking ties by rule order.

The input alphabet is left abstract: \ZipLex operates over sequences of an arbitrary type, allowing users to target, e.g., ASCII, UTF-8, or binary formats. To support invertibility, rules associate regular expressions with injective transformations from substrings to semantic values (to build semantic tokens\cite{egolf2022verbatim++}), together with their inverses. This guarantees that printing a token reproduces exactly the substring it was created from. To be able to reason about invertibility, \ZipLex provides functions to print sequences of tokens to strings, the main one being $\print$. We defer the formal treatment of invertibility to Section~\ref{sec:invertibility}.

\smartparagraph{Longest match specification}
In addition to invertibility, the correctness criterion for \ZipLex is the maximal munch principle~\cite{cattell1978formalization}, or \textit{longest match semantics}. For each produced token, no rule matches a strictly longer prefix of the remaining input, and no higher-priority rule matches an equally long one. \ZipLex verifies this property in Stainless, which, by induction on the input, establishes longest match semantics for the entire token sequence.

\section{Invertibility}
\label{sec:invertibility}

This section formalizes invertibility of lexing. We differentiate the two directions:
\begin{itemize}
    \item $\print(\lex(s)) = s$, implying the injectivity of $\lex$;
    \item $\lex(\print(ts)) = ts$, implying the injectivity of $\print$.
\end{itemize}
We explore these two directions separately in Section~\ref{subsec:injectivity-of-lex} and Section~\ref{subsec:injectivity-of-print}, respectively. Our notation uses 0-based indexing $s(i)$ for sequences, and (following Scala) $s \concat t$ for concatenation. For readability, we use $[a\text{-}z]$ and $[0\text{-}9]$ for unions of characters, and $r^+$ for $r\cdot r^*$.

\subsection{Ensuring \texorpdfstring{$\print(\lex(s)) = s$}{print(lex(s))=s}}
\label{subsec:injectivity-of-lex}

\smartparagraph{Semantic values and injection}
% \label{subsec:semantic-values-injection}
To achieve invertibility, we first ensure that each token is printed as the same string that was consumed to produce it. This is trivial when tokens store substrings directly, but in practice tokens often store semantic values~\cite{egolf2022verbatim++}. A transformation from strings to semantic values is sufficient to construct tokens, but insufficient to print them back to the same strings. For this reason, \ZipLex requires each rule to contain an injective transformation $\transform$ together with a corresponding $\charsOf$ function and a proof that 
% \[
$\forall s.\ \charsOf(\transform(s)) = s.$
% \]
Surjectivity is not needed because tokens only contain values created by $\transform$. Users are free to pick types and transformations that fit subsequent token manipulation. If the desired transformation is not injective (or if proving injectivity is inconvenient), one can always fall back on $\transform(s)=(v,s)$ with $\charsOf((v,s))=s$. 

Our approach uses standard assume-guarantee reasoning for sequential programs with preconditions and postconditions to require that token transformations are injective as above. Stainless thus formally makes sure that clients of \ZipLex establish injectivity to be able to use the lexer. To streamline these proofs for library clients, \ZipLex provides an \lstinline{Injection} construct that captures the semi-inverse relation between $\transform$ and $\charsOf$.
%(definition in Appendix~\ref{app:injection}). 
The definition of injection relies on support of Stainless for explicitly instantiated quantifiers in specifications. These proofs are very localized and, in our experience, easy.

From the required injectivity property, \ZipLex proves $\print(\lex(s)) = s$ by induction on $s$: at each step, injectivity ensures that printing the first token reconstructs the consumed prefix, and the induction hypothesis concludes the proof on the suffix. 

\subsection{Towards \texorpdfstring{$\lex(\print(ts)) = ts$}{lex(print(ts))=ts}}
\label{subsec:injectivity-of-print}

Unlike the previous case, $\lex(\print(ts)) = ts$ is not true in general. For example, with a rule based on $[a\text{-}z]^+$, printing a list containing two tokens, $\tkn{foo}$ and $\tkn{bar}$ yields $\texttt{"foobar"}$, which is tokenized as $\tkn{foobar}$. More generally, collapse may depend on more than two adjacent tokens: with token rules based on $a^*$, $b^*$, and $abc$, the sequence \tkn{a}\ \tkn{b} can be printed safely, while \tkn{a}\ \tkn{b}\ \tkn{c} cannot. We call a token sequence \emph{separable} if it satisfies $\lex(\print(ts)) = ts$.

We consider two approaches to ensure separability: (1) a special class of \textit{separator} tokens using a disjoint set of characters from other classes, and (2) an approach based on \textit{R-Path} conditions. Although usable when designing serialization formats, the separator-token approach is restrictive for lexing of programming languages: for example, whitespaces appear naturally as separators but also appear in strings and comments. We therefore 
%relegate it to Appendix~\ref{subsec:invertibility-separator-tokens} and 
focus on the more expressive approach, (2).

\subsection{\textit{R-Path} Predicates}
\label{subsec:r-path-predicate}

As a simple basic abstraction, we introduce a family of predicates on sequences where a binary relation must hold for each pair of consecutive elements. We call such predicates \emph{R-Path conditions}. An example of an R-Path condition is the \textit{sorted} predicate where $R$ is a total order. More generally $R$-paths are contiguous fragments of traces of a transition system given by the relation $R$.
For a relation $R$, we thus define $p_R(s)$ as
$\forall i.\ 0 \leq i < |s|-1 \implies R(s_i, s_{i+1}).$

Checking an R-Path requires a number of evaluations of $R$ linear in the length of the sequence. When working with a deductive verifier, R-Paths are convenient to maintain as invariants: slices of R-Paths are R-Paths, and concatenation requires only a check at the boundary (where $n_1$ is the length of $s_1$):
\begin{align*}
    p_R(s) &\implies p_R(s.\texttt{slice}(i,j)),\ \ \textit{for}\ 0 \leq i \leq j < |s|,\\
    p_R(s_1 \concat s_2) &\iff p_R(s_1)\ \land\ R(s_1(n_1-1), s_2(0))\ \land\ p_R(s_2),
\end{align*}
% where $n_1$ is the length of $s_1$.

\subsection{R-Paths That Ensure Invertibility}
\label{subsec:invertibility-r-path-predicate}

We define printable sequences as R-Paths for a particular relation $R$. The intuition is that tokens $t_1$ and $t_2$ are separable iff the first character of $t_2$ is already sufficient to ensure that $t_1$ is the longest token that can match in the printed string (regardless of any subsequent tokens following $t_2$).
We thus take $R$ to be the relation $\sep(t_1,t_2)$:
\[
\sep(t_1,t_2) := \neg \Big(\bigcup_{i\in m} r_i\Big).\prefixMatch(t_1 \concat t_2(0)),
\]
where $r_i$ is the regular expression of rule $i$. For notational simplicity, we view tokens as strings. The predicate $\Big(\bigcup_{i\in m} r_i\Big).\prefixMatch(s)$ means that some rule matches a string with prefix $s$. Equivalently, $s \in \bigcup_{k=1}^m \prefixSet(r_k)$ where $\prefixSet(r)=\{\,s\mid \exists s'.\ (s \concat s')\in L(r)\,\}$.

We then define separability for a length-$n$ token sequence by the R-Path condition induced by $\sep$:
\[
\sep(t_1,t_2,\dots,t_n) \iff \forall i.\ 1 \leq i < n \implies \sep(t_i,t_{i+1}).
\]

The relation $\sep(t_1,t_2)$ is local with respect to tokens but not rules: it must consider all regular expressions appearing the lexer's set of the rules. A relation local to both tokens and their producing rules cannot be sufficient: given a sequence $ts$ satisfying such a hypothetical local predicate, one can add a new rule matching exactly $\print(ts)$, in which case lexing $\print(ts)$ yields a single token, while the local predicate remains satisfied.

We illustrate $\sep$ on an example. Consider $r_1=[a\text{-}z]^+$ and $r_2=[0\text{-}9]^+$. For $t_1=\tkn{foo}$ and $t_2=\tkn{bar}$, the prefix $\texttt{"foob"}$ lies in the prefix set of $[a\text{-}z]^+$, so $\sep(t_1,t_2)$ does not hold. For $t_1=\tkn{foo}$ and $t_2=\tkn{123}$, $\texttt{"foo1"}$ is in no rule prefix set, so the tokens are separable. Because no rule matches any string starting with $\texttt{"foo1"}$, extending the sequence on the right cannot create a match either.

The function $\prefixMatch$ follows the same algorithm as the regular expression matching function $\matchR$ (described in Section~\ref{subsec:regexp-definition-implementation}), but when reaching the end of the prefix it checks that the resulting language is non-empty rather than nullable. This enables $\prefixMatch$ to take advantage of the memoization of derivatives, using the same cache (Section~\ref{sec:memoization}). We evaluate the cost of checking the R-Path $\sep$ in Section~\ref{sec:performance}.

We provide an abstract type $\PrintableTokens$ that maintains the R-Path invariant for $R=\sep$ on a given sequence. The invariant is computed once at instantiation; slicing preserves it, and concatenation requires only a boundary check $\sep(t_{\mathit{last}},t_{\mathit{first}})$. 
%The interface and invariants are shown in Appendix~\ref{app:printabletokens-interface}.

Our definition is sound: it implies separability. At the same time, it is efficient to check. 
Whereas the condition could be made more permissive by taking more characters into account, our condition already provides most of the practical benefits at a low computational cost. A maximally permissive predicate corresponds to printing, re-lexing and checking the equality; it is correct but, from a practical point of view, unnecessarily inefficient when manipulating long token sequences.

\subsection{R-Path vs Separator Tokens}

We also prove that the token sequence implicitly produced by printing while inserting separator tokens 
%(Appendix~\ref{subsec:invertibility-separator-tokens}) 
always satisfies the R-Path condition $\sep(t_1,\dots,t_n)$. This shows that the R-Path predicate is more general than interleaving separator tokens, and that interleaving provides a simple way to obtain a separable sequence without any runtime checks  when a separator token is available.

\section{Regular Expressions Engine}
\label{sec:regexp-engine}

This section presents the fully verified regular expression matching engine underlying \ZipLex. We first describe a reference implementation based on na\"ive Brzozowski's derivatives with straightforward syntax trees for regular expressions. We then detail our development of the optimized variant using Huet's zippers as the regular expression representation.

\subsection{Definition and Implementation}
\label{subsec:regexp-definition-implementation}

Regular expressions are parameterized by an arbitrary alphabet type and include the usual constructs: empty string, empty language, single-character matching, union, concat, and Kleene star. In addition to these constructors, our regular expressions offer a character-set primitive with an explicitly instantiated set of alphabet elements. This allows to describe sets without instantiating and deriving deep unions. Strings are sequences of alphabet elements.

The matching function $\matchR(r,s)$ decides membership of $s$ in the language of $r$ using Brzozowski's derivatives~\cite{brzozowski1964derivatives, varatalu2025re}. The engine also provides $\findLongestMatch$, which computes the longest prefix of an input string matched by a regular expression. Rather than testing all prefixes, this function traverses the input once, computing at most one derivative per character and stopping early when the derived expression denotes the empty language. If no prefix matches, the empty string is returned. Lexing rules are restricted to non-nullable expressions, so the empty string is never a token.

\smartparagraph{Specification}
We encode the standard semantic specification of regular expression matching as a ghost function $\matchRSpec$ 
%(Appendix~\ref{subsec:matchRSpec}) 
and prove:
% \[
$\forall r,s.\ \matchR(r,s) = \matchRSpec(r,s).$
% \]

\subsection{Zipper Optimization}
\label{subsec:regex-zipper-optimization}

While conceptually simple, na\"ive derivative-based matching suffers from expression blow-up and redundant computations. To address this, we implement and verify an optimization based on an instance of Huet's zippers~\cite{huet1997zipper}, following the approach of Edelmann~\cite[Chapter~2]{edelmann_efficient_2021}.
Zippers provide an alternative representation of regular expressions as sets of \textit{context}s, where each \textit{context} is a sequence of regular expressions (Figure \ref{fig:zipper-adt-def-focus-unfocus-signature}). Intuitively, contexts represent concatenations of their regular expressions, while zippers represent unions of such contexts. This representation yields a more efficient derivative algorithm and is particularly well suited to memoization: in contrast to naive derivatives for regular expressions, the set of reachable zippers during derivation is provably finite~\cite[Chapter 2]{edelmann_efficient_2021}, substantially reducing cache misses and recovering key properties of on-the-fly transformation to a finite automaton.

\begin{figure}[bth]
    %\centering
\begin{lstlisting}
case class Context[C](exprs: List[Regex[C]]):
  require(exprs.forall(validRegex))
type Zipper[C] = Set[Context[C]]
def focus[C](r: Regex[C]): Zipper[C]
@ghost def unfocus[C](zl: List[Context[C]]): Regex[C]
\end{lstlisting}
    \caption{ADT definitions for zipper-based representation of regular expressions and signatures of \lstinline{focus} and \lstinline{unfocus} to transform regular expressions into zippers and back.}
    \Description{ADT definitions for zipper-based representation of regular expressions.}
    \label{fig:zipper-adt-def-focus-unfocus-signature}
\end{figure}

Figure~\ref{fig:zipper-adt-def-focus-unfocus-signature} shows the zipper definition and the interfaces for converting between regular expressions and zippers. The zipper matching function $\matchZ$ follows the same structure as $\matchR$, computing derivatives character by character and checking nullability.

\smartparagraph{Specification}
Zipper-based matching is a drop-in replacement for regular expression matching. It satisfies $\unfocus(z)=r \implies \matchR(r,s)=\matchZ(z,s)$
for all $s,r,z$. 
This theorem is fully verified using Stainless.

\section{Towards Acceptable Performance}
\label{sec:towards-acceptable-performance}

In this section, we describe the optimizations required to obtain acceptable performance while maintaining verifiability.

First, we implement tail-recursive versions of all functions recursing on input strings that are executed at runtime. Although recursive functions make inductive proofs and executable code similar and therefore easier to write, they lead to high stack usage and potential stack-overflow crashes on the Java Virtual Machine. While this does not necessarily impact performance in a perceptible way, it negatively impacts usability, as the program crashes for large inputs. In our experiments, this issue arises mainly with the \lstinline{size} operation on \lstinline{List}, with the main functions of the lexer, $\lex$ and $\print$, and with the $\findLongestMatch$ function (see Section~\ref{sec:memoization} for in-depth explanations). To address this problem, we write tail-recursive implementations and prove their equivalence to the original versions. This allows us to preserve the existing proofs while replacing the runtime implementations, using equivalence theorems to bridge the gap between proofs and execution.

Second, we replace \lstinline{List} by a more efficient and fully verified data structure named \lstinline{BalanceConc} to represent sequences of alphabet elements and tokens, in all code executed at runtime. This structure is specified by a \lstinline{List} and represents sequences as balanced binary trees with immutable arrays of up to parameter $N$ elements in the leaves, where $N$ is a compile-time constant. Proofs are agnostic of $N$ so changing its value does not affect the correctness proofs. In this work, we use $N=512$. It is inspired by the Balance Conc implementation of~\cite{MadhavanKuncak17Memoization}. For specification and proofs, however, we continue to rely on \lstinline{List}. Although \lstinline{List} does not provide satisfactory runtime performance, it is well suited for verification: the Stainless library provides many lemmas and properties for lists, and the SMT theory of algebraic data types enables effective reasoning about them within Stainless. Using \lstinline{BalanceConc} for execution and \lstinline{List} for specification thus yields a practical combination of efficiency and verifiability.

Third, we implement memoization to improve performance, as described in Section~\ref{sec:memoization}, both for zipper derivatives and for inner functions of the lexer.

\section{Verified Memoization}
\label{sec:memoization}

In this section, we present a fully verified and efficient memoization framework. The framework is backed by a verified hash table we developed based on the work of Chassot et al.~\cite{chassot2024verifying}. The framework is reusable, thanks to a small set of lemmas that simplify the implementation and verification of memoization caches. We then show how we apply this framework to \ZipLex to improve performance without sacrificing correctness.

As shown by Reps~\cite{DBLP:journals/toplas/Reps98}, longest match lexical analysis can be performed in linear time in the size of the input string using memoization. However, memoization is challenging in a formally verified setting: it requires a map-like data structure that is both efficient and fully verified. In proof assistants such as Rocq, previous works including \vbatim~\cite{egolf2022verbatim++} and a verified implementation of orthologic proof search~\cite{DBLP:conf/cav/GuilloudP25} rely on standard-library maps, such as finite maps based on lists of pairs\footnote{\tiny\url{https://github.com/rocq-prover/stdlib/blob/master/theories/FSets/FMapList.v}} or more efficient (but at best logarithmic) AVL-tree-based maps\footnote{\tiny\url{https://github.com/rocq-prover/stdlib/blob/master/theories/FSets/FMapAVL.v}}. Even with such data structures, adding memoization typically requires threading caches through functions in a monadic style, including in proofs, which significantly increases verification effort and limits reusability.

To address these issues, we build on the verified mutable hash table of Chassot et al.~\cite{chassot2024verifying}, which verifies Scala's \texttt{LongMap}\footnote{\tiny \url{https://github.com/scala/scala/blob/2.13.x/src/library/scala/collection/mutable/LongMap.scala}}, an optimized hash table indexed by 64-bit integers. This allows us to obtain efficient memoization without re-verifying low-level hash-table machinery.

\subsection{Memoization Framework}

We first define a verified hash table with arbitrary-type keys, \lstinline{HashMap[K,V]}, implemented as a decorator of the verified \lstinline{LongMap}. Values stored in the \lstinline{LongMap} are linked lists of key-value pairs, yielding an instance of type \lstinline{LongMap[List[(K,V)]]}. We use the hash of the keys to index in the \lstinline{LongMap}. Building on top of the already verified \lstinline{LongMap} allows us to support generic keys at moderate cost without sacrificing too much in efficiency, provided that hash collisions remain reasonable, as the \lstinline{LongMap} is one of the most efficient implementations of a hash table in the Scala standard library. This design yields an implementation and verification effort of approximately 1600 additional lines of code, compared to nearly 8000 lines reported by Chassot et al.~\cite{chassot2024verifying}. We also implement and verify a functional executable specification for the \lstinline{HashMap[K,V]}.

To support memoization, we provide a small set of lemmas that capture the invariant of a memoization cache: when memoizing a function $f\!:\!A\to B$ using a hash table $h$, then $\forall(k,v)\in h.\ v=f(k)$. The lemmas show that the invariant is preserved by cache operations such as insertion. Notably, inserting a pair $(k',f(k'))$ into a cache that already satisfies the invariant yields a new state of the cache that continues to satisfies the invariant.

Using these lemmas, implementing a memoization cache requires only a few tens of lines of code, largely consisting of forwarding operations to the underlying hash table and applying the invariant-preservation lemmas. The mutable nature of the cache avoids monadic threading: the cache can be passed through the program directly, and call sites simply check for cached values, computing and inserting results on cache misses. We view this as evidence of a usability advantage of a verifier that supports mutation.

\subsection{Applying Memoization to \ZipLex}

The memoization framework is general and applies to any pure function. In \ZipLex, we use it to memoize the two zipper-derivative functions, \lstinline{derivStepUpZ}, \lstinline{derivStepDownZ}, and the function computing the longest matching prefix.

\smartparagraph{Zipper derivatives}
Memoizing zipper-derivative functions follows the standard pattern described above: derivative results are cached and reused across calls, substantially reducing repeated computations.

\smartparagraph{Longest match prefix}
Memoizing the longest-match computation requires additional care. The naive $\findLongestMatch$ function has two drawbacks: it relies on plain recursion
%, which can cause stack overflows on the JVM, 
(see Section~\ref{sec:towards-acceptable-performance}) and it operates on progressively shorter input strings (due to the prefixes being consumed to create tokens), which reduces memoization opportunities. To address these issues, we implement a tail-recursive algorithm that operates on the full input string together with indices.

\begin{figure}[!ht]
    \centering
    \begin{lstlisting}
def findLongestMatch[C](z: Zipper[C], input: Seq[C], testP: Seq[C], suffix: Seq[C]): 
  (Seq[C], Seq[C]) = 
    require(input == testP ++ suffix)
    if z.lostCause then (Nil, input) // shortcut
    else if testP == input then 
      if z.nullable then (testP, Nil) else (Nil, testP)
    else
      val newP = testP ++ List(suffix.head)
      val rec = findLongestMatch(z.deriv(suffix.head), input, newP, suffix.tail)
      if z.nullable and rec._1.isEmpty then (testP, suffix) else rec
def findLongestMatchTR(z: Zipper[C], fullInput: Seq[C], 
    input: Seq[C]): (Seq[C], Seq[C]) = 
  require(fullInput.size == input)
  val from = fullInput.size - input.size
  val lastNullablePos = 
    furthestNullablePos(z, fullInput, from, if z.nullable then from-1 else -1)
  val prefixL = lastNullablePos - from + 1
  if prefixL < 0 then input.split(0) else input.split(prefixL)
def furthestNullablePos[C](z: Zipper[C], fullInput: Seq[C], 
    from: BigInt, lastNullablePos: BigInt): BigInt = 
  if from == fullInput.size || z.lostCause then lastNullablePos
  else 
    val derivZ = z.deriv(fullInput(from))
    val newLastNullable = if derivZ.nullable then from else lastNullablePos
    furthestNullablePos(derivZ, fullInput, from + 1, newLastNullable) // tailrec
case class StackFrame[C](z: Zipper[C], from: BigInt, lastNullablePos: BigInt, 
                         @ghost res: BigInt, @ghost totalInput: Seq[C])
type Stack[C] = List[StackFrame[C]]
def furthestNullablePosStackMem[C](z: Zipper[C], fullInput: Sequence[C], from: BigInt,
    lastNullablePos: BigInt, stack: Stack[C])(using ...): (BigInt, Stack[C])
    \end{lstlisting}
    \caption{Algorithms that find the longest prefix matching a given zipper and signatures of the memoized function used for $\findLongestMatch$.}
    \label{fig:find-longest-match-algorithms-and-memoized}
\end{figure}

Instead of returning the longest matching prefix directly, the new algorithm computes the position of the furthest nullable state reachable from a given index (with the function \lstinline{furthestNullablePosition}). This representation enables memoization, since all recursive calls operate on the same input string and differ only in their indices. Figure~\ref{fig:find-longest-match-algorithms-and-memoized} shows both the naive and the tail-recursive versions, $\findLongestMatch$ and $\findLongestMatchTR$ respectively.

Tail-recursive functions' memoization cannot be implemented directly using the technique mentioned earlier, since updating the cache would break tail recursion. We therefore maintain an explicit stack of encountered parameters. Once the tail-recursive computation completes, the calling function iterates over this stack and updates the cache for all recorded calls. Because all recursive calls return the same result, this strategy preserves correctness while enabling memoization. This can be seen as an explicit simulation of the JVM call stack in the heap, inspired by techniques used in continuation-passing style and related defunctionalization transformations.

The memoization of \lstinline{furthestNullablePosition} is equivalent to the technique proposed by Reps~\cite{DBLP:journals/toplas/Reps98}. As a result, lexing runs in $O(n)$ time, where $n$ is the input string length. To the best of our knowledge, this is the first verified implementation of longest match lexing with linear-time complexity. By contrast, \vbatim~\cite{egolf2022verbatim++} also verifies memoization, but due to its choice of data structures achieves $O(n\log n)$ complexity. We evaluate the performance impact in Section~\ref{sec:performance}.

As proposed by Reps~\cite{DBLP:journals/toplas/Reps98}, memory usage can be traded off against constant-factor performance by memoizing only a fraction of calls. We implement and verify this heuristic by storing only one out of every $k$ calls when the input string is longer than $M$ characters, where both $k$ and $M$ are computed by pure and opaque functions that can depend on the rules and the input string. The correctness proof is agnostic to these functions, so changing them does not affect verification. By default, we set them to the constant values $k=10$ and $M=10^6$.

% table before the title for more layout flexibility
\begin{table}[bth]
\centering
\caption{Line of code count per component}
\begin{tabular}{|l|r|r|r|}
\hline
Module & Program LOC & Spec + Proof LOC & Total \\
\hline
Longest Match Lexing & 580 &  2030 & 2610 \\
Lexer Printing + Invertibility & 130 & 1430 & 1560 \\
Regex Derivative Based Matching & 470  & 2440 & 2910 \\
Zipper Based Matching & 390 & 2670 & 3060 \\
BalanceConc & 255 & 320 & 575 \\
List \& Set lemmas & - & 1610 & 1610 \\
Mutable HashMap (excluding LongMap) & 90 & 2700 & 2790 \\
\hline
Total & 1'915 & 13'200 & 15'115 \\ 
\hline
\end{tabular}
\label{tab:loc}
\end{table}

\section{Verification Experience}

We discuss our experience formally verifying \ZipLex. In Stainless, proofs are written as Scala functions and expressions alongside executable code. Table \ref{tab:loc} shows the sizes of implementation and proof components in terms of lines of code. Overall, the proof-related code amounts to roughly ten times more lines than the executable implementation. We spent most of the time exploring strategies for inductive proofs. Aside from simple completions in VS Code, we did not rely on LLMs. In this section, we outline how each component (lexer, regular-expression engine, zipper optimization, and memoization framework) is specified, which induction schemes and ghost arguments are used to discharge proof obligations, and where the main proof challenges arise.

% \smartparagraph{General proof methodology}
% During the entire project, we follow this methodology: we first implement an inefficient but easier to verify version of the given algorithm, and then gradually verify more efficient implementations or optimizations, proving the equivalence with the previous step. For the regular expression matching, we started with plain matching (our executable specification), followed by derivatives, then zipper-based matching, then augmented with memoization. The lexer starts as a recursive and naive implementation, followed by a tail-recursive version, augmented with memoization. The same applies to the underlying data structure that starts as a plain singly linked list to then move to the more efficient \textit{BalanceConc}.
% This approach allows us to obtain very early a fully verified and executable implementation, that is then gradually made more efficient by some optimizations.
\smartparagraph{General proof methodology}
Over the course of development, we first implemented an intentionally inefficient but conceptually simple and easily verifiable version of the target algorithm. We then incrementally verified more efficient implementations or optimizations, each time proving their equivalence with the previous version.
For regular expression matching, we started from a straightforward matching procedure (our executable specification), then introduced Brzozowski derivatives, followed by a zipper-based algorithm, and finally added memoization. The lexer was first implemented as a naive recursive function, then transformed into a tail-recursive version, and finally extended with memoization. We applied an analogous refinement to the underlying data structure: we moved from a simple singly linked list to the more efficient \textit{BalanceConc} concatenable tree representation with arrays as described in Section~\ref{sec:towards-acceptable-performance}.
%We applied an analogous refinement to the underlying data structure: we moved from a simple singly linked list to the more efficient \textit{BalanceConc} concatenable tree representation based on \cite{DBLP:conf/lcpc/ProkopecO15} but simplified and verified to conform to a specification in terms of List.
%\footnote{\url{https://github.com/epfl-lara/bolts/}}. 
(We denote this data structure simply as \lstinline|Sequence|.)
This methodology yields a fully verified, executable implementation early on, which we then optimize for efficiency while preserving formal correctness.

\smartparagraph{Longest match}
We begin with the proof of the longest match semantics. We first establish this property for the list-based function \lstinline{lexList}, which uses plain recursion and relies on the intermediate functions \lstinline{maxPrefixOneRule} and \lstinline{maxPrefix} to compute, respectively, the longest prefix for each rule and the global longest prefix. The proof proceeds by induction on the input list and relies on two key ingredients: (1) the correctness of regular expression matching, in particular that \lstinline{findLongestMatch} is sound and returns the longest matching prefix, and (2) auxiliary lemmas establishing the soundness of \lstinline{maxPrefixOneRule} and \lstinline{maxPrefix}.

From this basis, we first show observational equivalence between the recursive \lstinline{Sequence}-based implementation \lstinline{lexRec} and \lstinline{lexList}, using the bridging specification of \lstinline{Sequence} as a \lstinline{List}. We then prove that the tail-recursive implementation \lstinline{lexTailRec} produces exactly the same output as \lstinline{lexRec} on all inputs. This final step requires helper lemmas about associativity of list concatenation and properties of \lstinline{lex}, such as the fact that removing a suffix of the input does not affect the lexing of the prefix, provided that the cut occurs at token boundaries. 
%The lexer implementation without printing and invertibility consists of about 490 executable lines of code, while the corresponding correctness proof amounts to approximately 1900 LOC.

\smartparagraph{Invertibility} We first establish invertibility for the list-based implementations of printing functions, and then prove equivalence with their sequence-based and tail-recursive counterparts. The proof of $\print(\lex(s)) = s$ proceeds by induction on $s$. For the converse direction, $\lex(\print(ts)) = ts$, we likewise perform induction on $ts$, supported by several auxiliary lemmas. These include, for example, that the separability predicate $sep(ts)$ ensures that the first token produced by $\lex(\print(ts))$ is $ts.\head$, and that the same property holds for sequences interleaving separator and non-separator tokens.
Several lemmas rely on properties of regular expressions, such as character sets used by the rules and their impact on prefix. 
%The implementation and proofs for lexer invertibility amount to approximately 1550 LOC, of which about 130 correspond to executable code.

\smartparagraph{Naive derivative-based regular expression matcher}
For the derivative-based regular expression engine, we prove correctness by handling each constructor case separately and then combining the resulting lemmas to establish the main theorem
% \[
$\forall r,s.\ \matchR(r, s) = \matchRSpec(r, s)$. 
% \]
For example, one lemma shows that if $r_1 + r_2$ matches a string $s$, then either $r_1$ or $r_2$ matches $s$. Conversely, another lemma proves that if $r$ matches $s$, then both $r + r'$ and $r' + r$ match $s$. Taken together, these lemmas establish the correctness of Brzozowski's derivative-based matching algorithm. 
%The executable implementation of this engine consists of about 435 lines of code, while the proofs amount to approximately 2815 LOC.

\smartparagraph{Zipper regular expression matching}
As explained in Section~\ref{subsec:regex-zipper-optimization}, the main correctness theorem for zipper-based matching is
% \[
$\forall s, z.\ \matchR(\unfocus(z), s) = \matchZ(z, s).$
% \]
We state and prove this theorem using a list-based representation of zippers (rather than sets) in order to enable induction on the list of contexts.

Although the derivative operations on regular expressions and on zippers are structurally similar and intuitively equivalent, the proof of this theorem is complex and requires approximately 600 LOC on its own. The main difficulty stems from the non-injectivity of the \lstinline{unfocus} function. An intuitive formulation of the equivalence would be
% \[
$\forall r, s.\ \matchR(r, s) = \matchZ(\focus(r), s),$
% \]
which suggests induction on the regular expression and the input string. However, this approach fails in our setting. After rewriting $\matchR$ using its specification, one would need properties such as
% \[
$\focus(\Union(r_1, r_2)) = \focus(r_1) \cup \focus(r_2)$
% \]
to apply the induction hypothesis. This equality does not hold for our implementation of \lstinline{focus}, even though unfocusing both sides yields equivalent regular expressions. As a result, the induction breaks down.

We therefore use the formulation
% \[
$\forall z, s.\ \matchR(\unfocus(z), s) = \matchZ(z, s)$, 
% \]
which allows induction on the zipper and the input string instead. This makes the proof possible, but significantly more involved, as \lstinline{unfocus} itself introduces an additional induction over the list of contexts.

In summary, despite the similar high-level structure of the two derivative algorithms, the fundamental difference between regular expressions and zippers complicates the proof. Regular expressions form tree-like structures with a clear recursive schema, while zippers are flat collections of contexts, which do not admit the same induction patterns. 
%The zipper-based implementation consists of about 370 LOC, while the corresponding proofs amount to approximately 2580 LOC.

\smartparagraph{Memoization framework}
As discussed in Section~\ref{sec:memoization}, the memoization framework relies on a verified mutable \lstinline{HashMap[K, V]}. To prove its correctness, we follow the approach of Chassot et al.~\cite{chassot2024verifying} by establishing equivalence between each mutable operation of \lstinline{HashMap[K, V]} and the corresponding operation on an immutable executable specification \lstinline{ListMap[K, V]} implemented as an immutable association list of pairs. We therefore first implement and verify a generic \lstinline{ListMap[K, V]}, similar to their \lstinline{ListLongMap[V]}. We then implement and prove the correctness of the additional layer that stores buckets as immutable linked lists of key-value pairs inside an underlying \lstinline{LongMap[V]}. For this proof, we rely on the correctness of the LongMap~\cite{chassot2024verifying}. The proofs of the memoization helper lemmas proceed by induction on the key-value pairs stored in the map.
%The mutable \lstinline{HashMap[K, V]} implementation consists of about 90 LOC, while the executable specification and proofs amount to approximately 2700 LOC.

\smartparagraph{Verification condition statistics}
Stainless generates 21'961 verification conditions (VCs) for the entire project (8'802 of which are cache hits), and takes $\approx$130 minutes to verify on the server described in Section \ref{sec:performance}, creating multiple GB of SMT-LIB files. We use the Stainless version from the commit at the time of paper preparation\footnote{\tiny\url{https://github.com/epfl-lara/stainless/tree/d89c8eccaf0261f8be348bee7ffa6f5a3c1c5ca6}}. VC verification times resemble a negative exponential distribution, with only 44 VCs taking more than 10 seconds to verify. 
%(Appendix Fig. \ref{fig:vcs_time_distribution}). 
The slowest VC takes 58 seconds to verify during the run with multiple approximating SMT-LIB queries. Stainless uses function unfolding, so it queries SMT solvers multiple times for a single VC. The final SMT-LIB file for this slowest query has 100 lines; cvc5 takes less than a second to re-verify it.

\section{Performance Evaluation}
\label{sec:performance}

% We present an evaluation of the lexer and the regular expression matching engine, highlighting the impact of the proposed optimizations.
To evaluate the practicality of \ZipLex, we present a performance evaluation investigating the following questions:
\begin{enumerate} \itemsep=0pt
    \item [Q1:] What complexity does \ZipLex exhibit on an adversarial grammar, and how does it compare to other state-of-the-art lexers?
    \item [Q2:] What is the cost of computing $\sep$ and what is the impact of the mix of static and runtime checks using the $\PrintableTokens$ constructs?
    \item [Q3:] What is the performance of \ZipLex on a realistic JSON grammar and how does it compare to other verified and unverified lexers?
\end{enumerate}
All experiments are run on a server running Linux with two sockets of an \textit{AMD EPYC 9254 24-Core} processor (base frequency 2.9\,GHz, up to 4.15\,GHz), offering 96 virtual processors and 128\,GB of RAM. Our algorithms are sequential and do not exploit parallelism. Error bars show 99.9\% confidence intervals.
We use the JMH benchmarking library \cite{jmhWebSite} from OpenJDK  to benchmark Scala code. We use OpenJDK JVM version \textit{21.0.2} unless stated otherwise.

\smartparagraph{Evaluating linear-time behavior via lexing $a$ and $a^*b$}
%\label{subsec:performance-lexing-a-astarb}
To evaluate the impact of memoization on lexing performance, we first consider a grammar with two rules, $r_1 = a$ and $r_2 = a^*b$, as in~\cite{Li_Mamouras_2025}. This grammar is adversarial for naive longest-match lexers: when the input string contains only '$a$'s, computing the longest match for $r_2$ requires consuming the entire remaining input to ensure that no '$b$' occurs at the end. As a result, naive lexing exhibits quadratic time complexity.
Although such grammars rarely appear in compiler lexers, they are relevant in the context of uniform tokenization, where rules are part of the input. In such settings, adversarial grammars can be used to mount denial-of-service attacks on lexers. Even if rare, such patterns would arise in a lexer having one rule for digits, $r_1 = digit$, and one for real number constants, $r_2 = digit^* \mbox{\tt .} digit$.

\begin{figure}
    \centering
    \begin{minipage}[t]{0.49\textwidth}
        \centering
        \includegraphics[width=\linewidth,alt={Scatter plot showing lexing time for different input sizes for the lexers ZipLex, Flex, Coqlex, and \vbatim. X-axis shows the input size in characters and ranges from 0 to 50'000; y-axis shows the time in seconds and ranges from 0 to 50. For 50'000 characters, Coqlex shows 50 seconds, Flex around 8 seconds, ZipLex and \vbatim are too close to 0 to see.}]{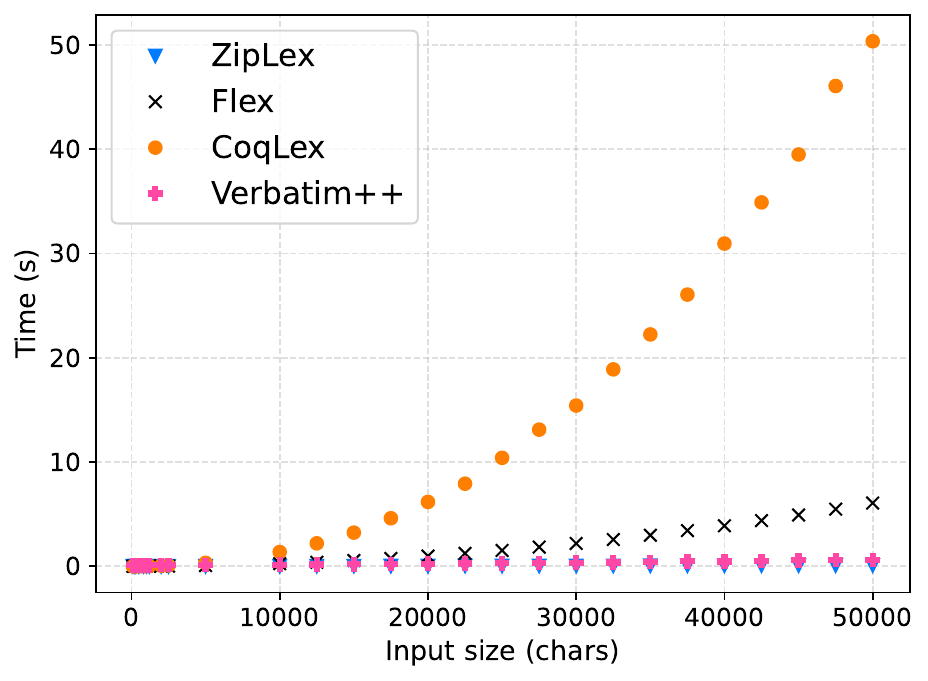}
    \end{minipage}
    \hfill
    \begin{minipage}[t]{0.49\textwidth}
        \centering
        \includegraphics[width=\linewidth,alt={Scatter plot showing lexing time for different input sizes for the lexers ZipLex, and \vbatim. X-axis shows the input size in characters and ranges from 0 to 100'000; y-axis shows the time in seconds and ranges from 0 to 1.2. \vbatim shows data from x=0 up to x=85'000; for x=85'000, y=1.2 seconds. ZipLex shows data from x=0 to x=100'000; for x=100'000, y=0.2 seconds.}]{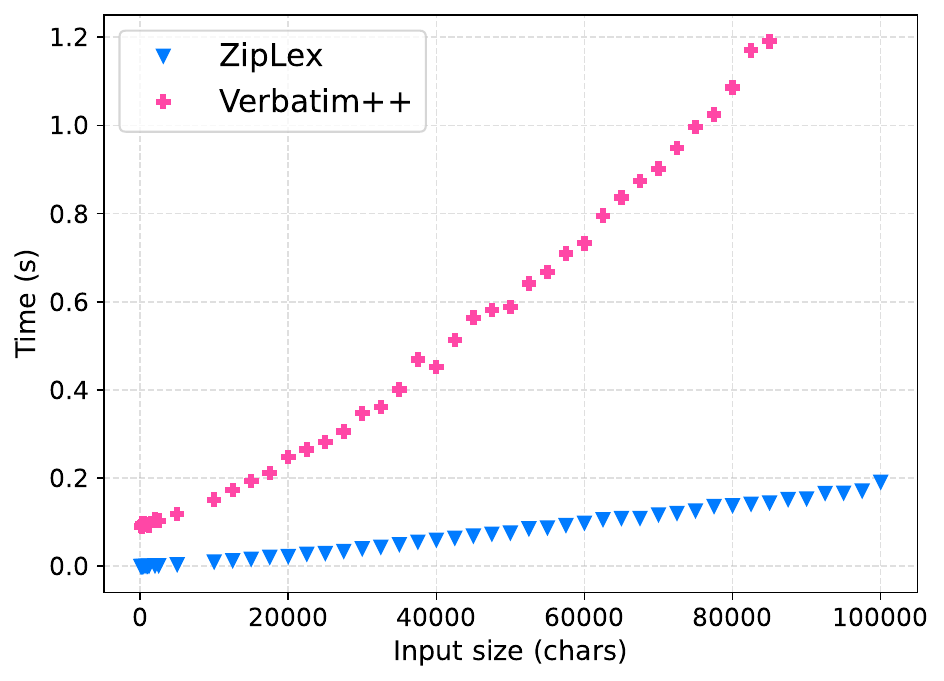}
    \end{minipage}

    \caption{Lexing time with grammar rules $r_1=a$ and $r_2=a^*b$, on strings of '$a$'s of sizes ranging from 100 to 50K (or 100K) comparing  \ZipLex, flex, Coqlex, and \vbatim.} 
    \label{fig:a-a-star-b-lexing-lexers-comparisons}
\end{figure}

\begin{figure}[!htb]
    \centering
    \begin{minipage}[t]{0.49\textwidth}
        \centering
        \includegraphics[width=\linewidth,alt={Scatter plot showing lexing time for different input sizes for different lexers (see caption). X-axis shows the input size in characters and ranges from 0 to 60'000; y-axis shows the time in seconds and ranges from 0 to 8. For 50'000 characters, all \vbatim variants show around 6 to 8 seconds, while the other lexers are too close to 0 to properly identify.}]{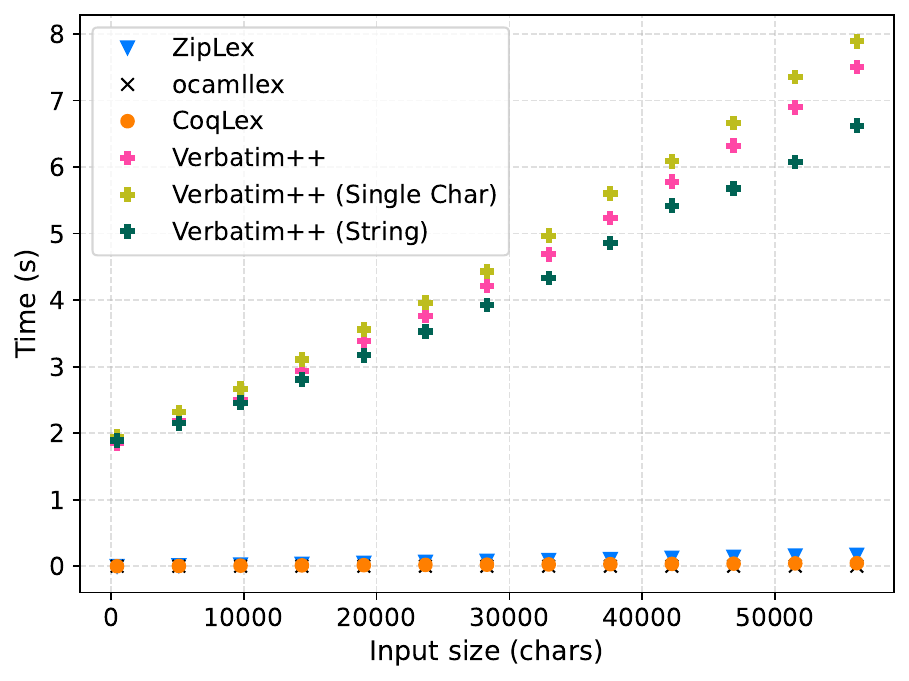}
    \end{minipage}
    \hfill
    % \begin{minipage}[t]{0.48\textwidth}
    %     \centering
    %     \includegraphics[width=\linewidth]{res/JSON_Lexing_CoqLex_vs_ZipLex_vs_Ocamllex.pdf}
    % \end{minipage}
    \begin{minipage}[t]{0.49\textwidth}
        \centering
        \includegraphics[width=\linewidth,alt={Scatter plot with confidence interval and a linear regression as a line. X-axis shows the input size in thousands of characters and ranges from 0 to 30'000; y-axis shows the time in seconds and ranges from 0 to 300. The linear regression fits the data. For x=30 millions characters, y=250 seconds, with the confidence interval [200, 400].}]{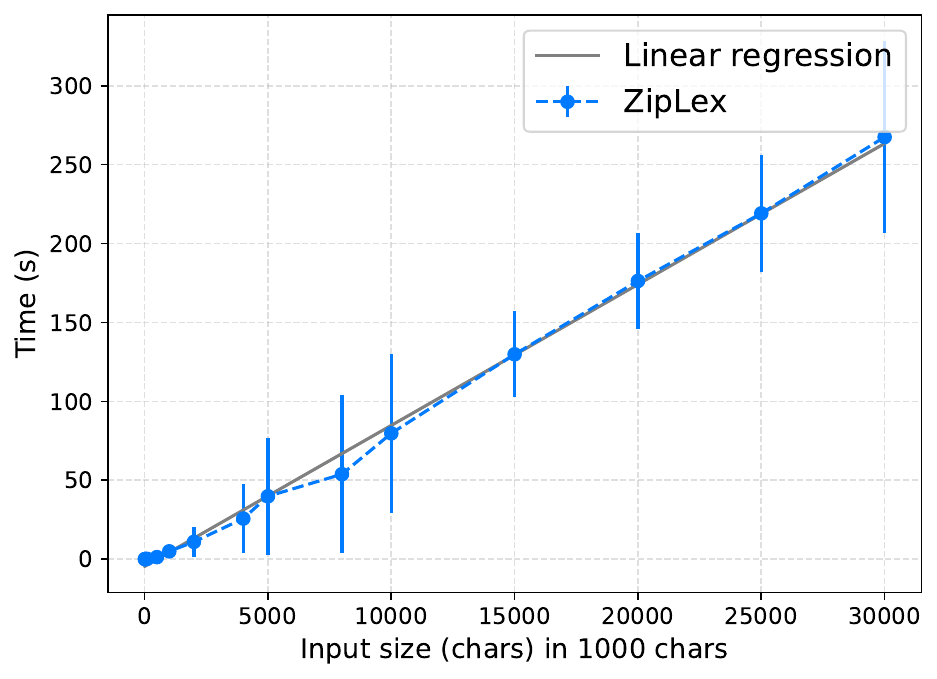}
    \end{minipage}

    \caption{Left: JSON lexers performance comparison; x-axis shows the length of input files in number of characters; y-axis shows the time in seconds to lex the input file; comparison between \ZipLex, Coqlex, OCamllex, and variants of \vbatim. Right: \ZipLex lexing time with grammar $r_1=a$ and $r_2=a^*b$, on strings of $a$s of sizes ranging from 10K to 30M of characters, with a linear regression of degree 1.}
    \label{fig:lexer-json-comparisons-baselines-and-a-a-star-b-lexing-ziplex-regression}
\end{figure}

Figure~\ref{fig:a-a-star-b-lexing-lexers-comparisons} compares lexing times for strings of '$a$'s up to 50\,000 characters using flex~\cite{paxson1995flex}, Coqlex~\cite{ouedraogo2023coqlex}, \vbatim~\cite{egolf2022verbatim++}, and \ZipLex. Both flex and Coqlex exhibit quadratic behavior on this grammar.
%, while \vbatim is super-linear. 
\vbatim stack overflows beyond 85\,000 characters.
Figure~\ref{fig:lexer-json-comparisons-baselines-and-a-a-star-b-lexing-ziplex-regression} (Right) shows \ZipLex's performance in the same setting (with larger files), experimentally confirming the linear complexity.

\begin{figure}[bth]
    \centering
    \begin{minipage}[t]{0.49\textwidth}
        \centering
        \includegraphics[width=\linewidth,alt={Scatter plot: x-axis shows the input size in characters and ranges from 0 to 1'000'000; y-axis shows the time in seconds and ranges from 0 to 6. The points aligned on 2 lines. For x=1'000'000, y=5.5 for lexing and computing the R-Path condition and y=5 for lexing. Both confidence intervals overlap.}]{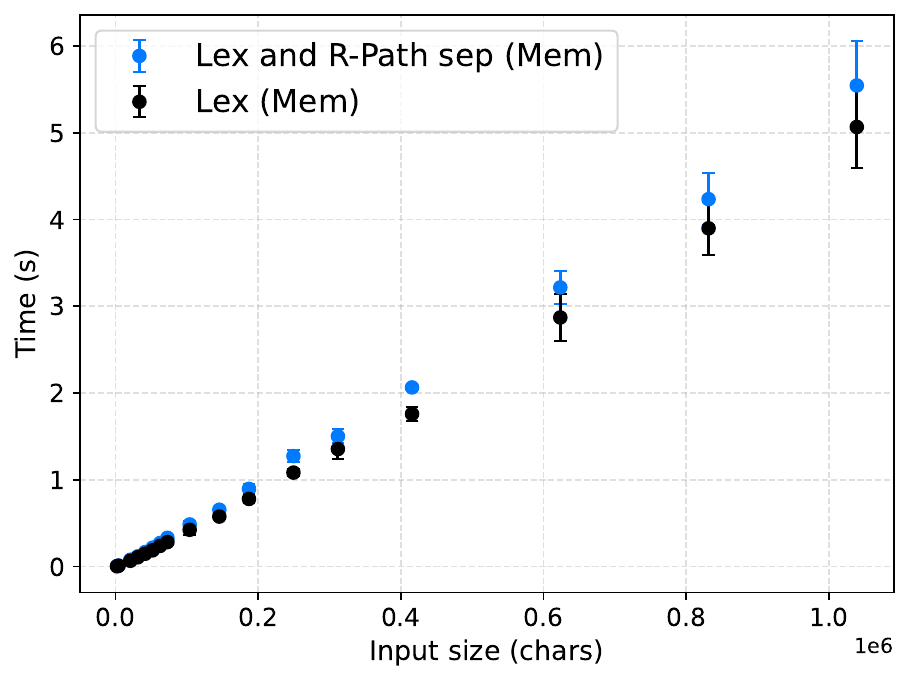}
    \end{minipage}
    \hfill
    \begin{minipage}[t]{0.49\textwidth}
        \centering
        \includegraphics[width=\linewidth,alt={Scatter plot: x-axis shows the input size in characters and ranges from 0 to 1'000'000; y-axis shows the time in miliseconds and ranges from 0 to 800. The points aligned on 2 lines. For x=1'000'000, y= around 600 for concatenating with the wrapper, with a confidence interval [390, 800]. For concatenating with the wrapper, y is too close to zero to properly identify.}]{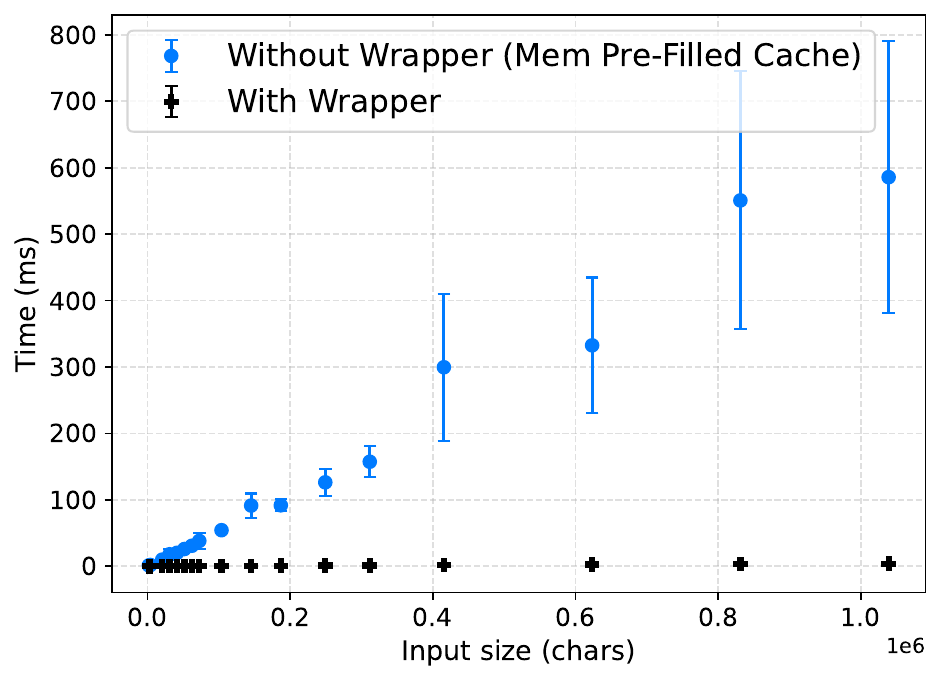}
    \end{minipage}

    \caption{\textbf{Left:} Comparison between lexing and computing R-Path \texttt{sep} on the sequence of tokens (instantiating \lstinline{PrintableTokens} with the obtained tokens) against lexing only, both using memoization. \textbf{Right}: Comparison between recombining slices as instances of \lstinline{PrintableTokens} and creating a sequence and computing the R-Path \texttt{sep}.} 
    \Description{Left: Comparison between lexing and computing R-Path \texttt{sep} on the sequence of tokens (instantiating \lstinline{PrintableTokens} with the obtained tokens) against lexing only, both using memoization. Right: Comparison between recombining slices as instances of \lstinline{PrintableTokens} and creating a sequence and computing the R-Path \texttt{sep}.} 
    \label{fig:json-manipulation-R-path-comparison-combination-with-without}
\end{figure}

\smartparagraph{Evaluating printable sequences on JSON lexing and sorting}
We next evaluate \ZipLex on the JSON object-sorting application presented in Section~\ref{sec:usage-examples}. We focus on the cost of enforcing separability through the R-Path predicate.
Figure~\ref{fig:json-manipulation-R-path-comparison-combination-with-without} (left) compares lexing alone with lexing followed by instantiating a \lstinline{PrintableTokens} wrapper, which computes the R-Path predicate $\sep$. In both cases, memoization is enabled. The results show that the overhead of computing $\sep$ is small. This is partly explained by the fact that the derivative cache is already populated during lexing. In practice, we expect \lstinline{PrintableTokens} to be instantiated immediately after lexing, reusing the cache and keeping the overhead low.

Using \lstinline{PrintableTokens} provides a significant performance benefit when slicing and recombining token sequences. Figure~\ref{fig:json-manipulation-R-path-comparison-combination-with-without} (right) compares recombining slices represented as \lstinline{PrintableTokens} with recombining plain token sequences followed by recomputing the $\sep$. Concatenating two \lstinline{PrintableTokens} instances requires only a constant-time check of $\sep(t_1,t_2)$ at the boundary (Section~\ref{subsec:invertibility-r-path-predicate}), since $\sep$ is maintained as a statically verified invariant. Consequently, recombining slices is linear in the number of \emph{slices}, whereas recomputing $\sep$ is linear in the number of \emph{tokens}. 
%When the number of slices is small, recombination incurs almost no overhead, as shown in the plot.
%
These results illustrate that combining static guarantees with lightweight runtime checks, as done by \lstinline{PrintableTokens}, provides an effective way to obtain verified and efficient programs.
The right plot of Figure~\ref{fig:json-manipulation-R-path-comparison-combination-with-without} also shows that concatenation of large \lstinline{Sequence}s is efficient, confirming the suitability of this data structure for runtime use.

\begin{figure}[!ht]
    \centering
    \hspace*{-3mm}
     \begin{minipage}[t]{0.54\textwidth}
        \centering
        \includegraphics[width=\linewidth,alt={Scatter plot: x-axis shows the input size in number of characters and ranges from 0 to 60'000; y-axis shows the time in seconds and ranges from 0 to 0.175. For x=60'000, ZipLex shows around y=0.175, ZipLex with memoization of derivatives shows around y=0.125, ZipLex with no memoization shows around y=0.1, Coqlex shows around y=0.05, and OCamlLex is too close to 0 to properly identify.}]{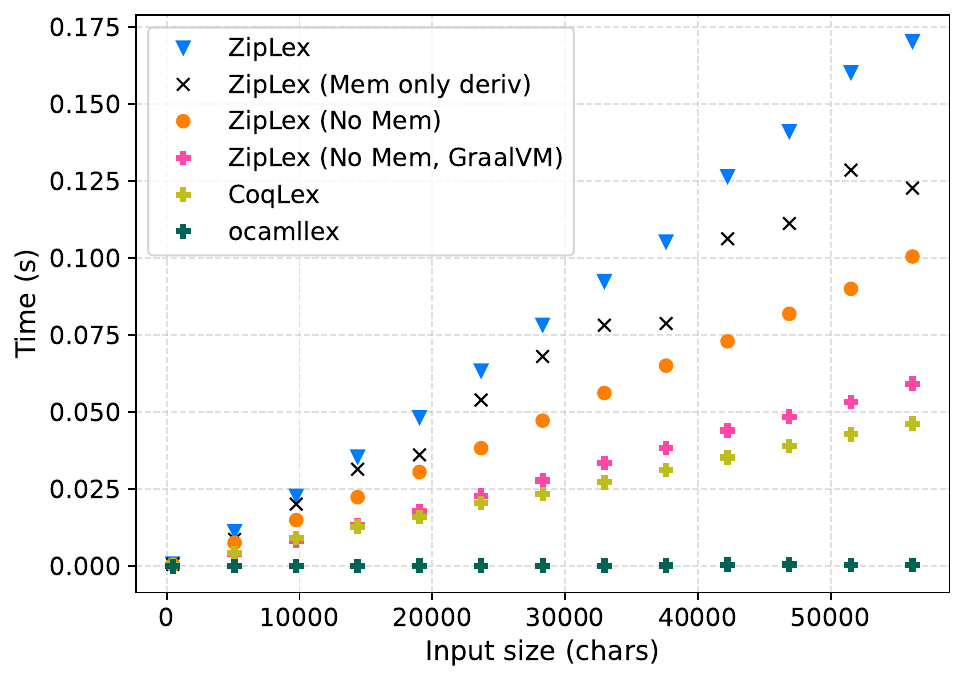}
    \end{minipage}
    \hspace*{-3mm}
    \begin{minipage}[t]{0.45\textwidth}
        \centering
        \includegraphics[width=\linewidth,alt={Scatter plot: x-axis shows the input size in number of entries and ranges from 0 to 9'000'000; y-axis shows the time in seconds and ranges from 0 to 14. For x=9'000'000, the VerifiedLongMap shows around y=2, the Scala HashMap shows around y=4, the verified HashMap shows around y=8, and the Scala immutable HashMap shows around y=12 seconds.}]{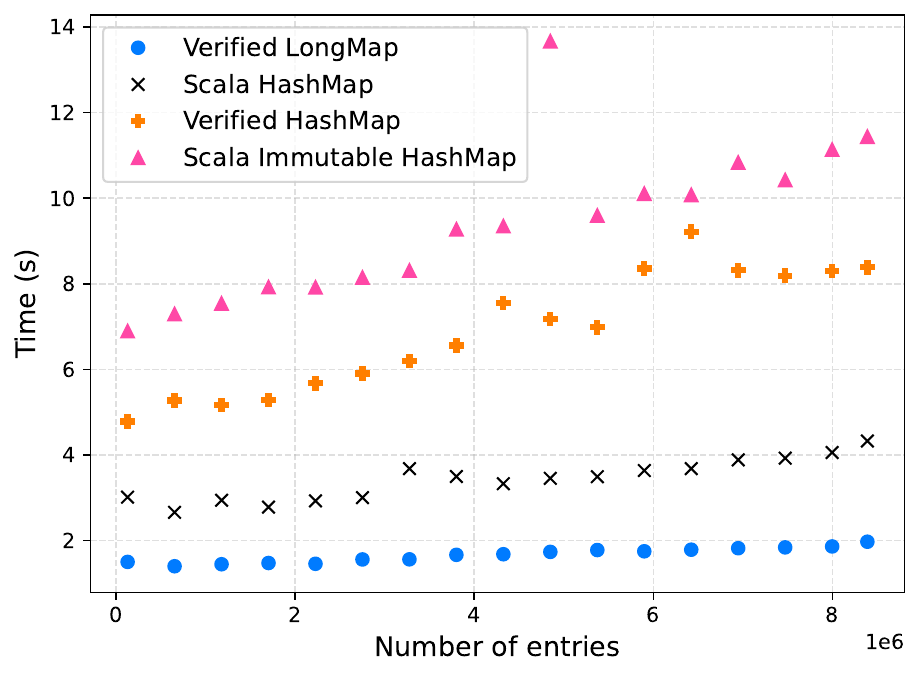}
    \end{minipage}

    \caption{\textbf{Left}: JSON lexers performance comparison; x-axis shows the length of input files in number of characters; y-axis shows the time in seconds to lex the input file; comparison between Coqlex, OCamllex, and different memoization variants of \ZipLex, all memoization (named ZipLex), only derivative computations memoized (named \textsf{ZipLex (Mem only deriv)}), no memoization (named \textsf{ZipLex (No Mem)}), and \textsf{ZipLex (No Mem)} run on Oracle GraalVM. \vbatim is not shown as it is several orders of magnitude slower, see Figure~\ref{fig:lexer-json-comparisons-baselines-and-a-a-star-b-lexing-ziplex-regression}.
    \textbf{Right}: Performance evaluation of Scala library immutable map, Scala library mutable hashmap, the verified LongMap hash table, and our verified hash map on a benchmark performing the following tasks: insert N pairs into the map, remove half of them, and update the N pairs again.}
    \label{fig:hashmap_performance_comparison_json_ziplex_variants}
\end{figure}

\smartparagraph{Performance comparison of JSON lexers}
% \label{subsec:performance-lexers-competitions}
We compare \ZipLex against verified and unverified JSON lexers using the benchmark inputs from the Coqlex evaluation~\cite{ouedraogo2023coqlex}. We reuse the benchmarking infrastructure released with Coqlex~\cite{coqlex_comparisonjson_2023_repo}, combined with our own infrastructure, to compare \ZipLex with Coqlex~\cite{ouedraogo2023coqlex}, \vbatim~\cite{egolf2022verbatim++}, OCamllex~\cite{Ocamllex}, and two additional \vbatim variants from Ouedraogo et al.~\cite{ouedraogo2023coqlex}. All lexers are formally verified except for OCamllex (but only \ZipLex verifies invertibility). The benchmark consists in lexing each file and measuring total execution time.
Figure~\ref{fig:lexer-json-comparisons-baselines-and-a-a-star-b-lexing-ziplex-regression} (left) shows the results. Unlike \vbatim, which relies on a transformation from regular expressions to DFAs and incurs a fixed preprocessing overhead of approximately two seconds, \ZipLex's zipper-based matching and memoization introduce no such upfront cost. \ZipLex is approximately $3.5\times$ slower than Coqlex on JSON inputs, while remaining two orders of magnitude faster than \vbatim. To better assess the reasons behind the difference with Coqlex, we run the same experiments with two new variants of \ZipLex with reduced memoization: one with memoization enabled only for the zipper-based derivative computations, and one with no memoization enabled. We also run our experiments using the Oracle GraalVM~\cite{DBLP:conf/pldi/WurthingerWHWSS17} (version \textit{21.0.11}) instead of OpenJDK. The results are shown in Figure~\ref{fig:hashmap_performance_comparison_json_ziplex_variants} (left). 
The results show that, while memoization ensures linear complexity for lexing under all circumstances, it is detrimental to absolute performance in the particular case of the JSON grammar. Indeed, without any memoization enabled, \ZipLex is faster when lexing JSON, and the variant with memoization of derivatives lies between \ZipLex with full memoization and \ZipLex with no memoization. Running on GraalVM without any memoization, \ZipLex is faster than on OpenJDK, and is $1.37\times$ slower than Coqlex. Due to differences in the underlying runtimes (OCaml vs different JVMs), we are unable to definitely establish the source of the remaining 1.37$\times$ performance gap. 
We suspect that a key difference is that Coqlex lexers are specialized for the alphabet of ASCII characters and can make use of character intervals, whereas \ZipLex characters are entirely generic, leading to boxing and explicit sets of characters in the zipper data structure.

We believe that \ZipLex's performance is acceptable given its design goals. In addition to  correctness guarantees with respect to regular expressions and longest match semantics, \ZipLex provides invertible lexing and supports arbitrary alphabets, offering greater flexibility than existing verified lexers, as well as guaranteed linear behavior in the input length thanks to verified memoization.
%, independently of the specific grammar rules.

\smartparagraph{Performance of the mutable hash table}
Our verified hash table, being mutable, allows us to add memoization to functions with less refactoring that needed with an immutable cache. One could implement a mutable cache as a variable containing an immutable map which would lead to a smaller verification effort. Such immutable structure would, however, lead to worse performance, as shown in Figure~\ref{fig:hashmap_performance_comparison_json_ziplex_variants} (right). The figure measures time for the following operations: insert N random pairs to the map under evaluation, remove half of those pairs, then insert all N pairs again. We compare our verified hash map, verified LongMap\cite{chassot2024verifying}, Scala's standard library's mutable hash map, and its immutable HashMap. Our mutable HashMap is slower than Scala library's unverified hashmap but outperforms Scala library's immutable HashMap, especially for large numbers of pairs. This demonstrates that using a mutable hash table for caches improves performance and justifies the verification effort.

\section{Conclusion}

\ZipLex is a verified framework for invertible lexical analysis with linear worst-case complexity. 
Unlike past verified lexers, which only ensure the semantics of matching regular expressions and maximal munch, 
\ZipLex also guarantees that lexing and printing are mutual inverses. 
Our design relies on two key ideas: (1) a new class of predicates, 
\emph{R-Paths}, that capture the separability of token sequences, and 
(2) a combination of verified data structures and optimizations, 
including Huet's zippers and memoized derivatives and lexing, to achieve practical matching performance.
We implemented \ZipLex in Scala and verified its correctness and invertibility using the Stainless verifier. 
Our evaluation demonstrates that \ZipLex supports applications such as 
JSON processing and compiler lexers. Thanks to verified memoization based on a verified mutable hash table, \ZipLex exhibits linear complexity in the worst case, even on grammars for which state-of-the-art lexers such as flex and OCamlex behave quadratically. On JSON lexing, \ZipLex is a factor of $3\times$ slower than the best verified quadratic lexer, Coqlex, but two orders of magnitude faster than \vbatim. By disabling memoization, the performance gap with Coqlex can be reduced to a factor of $1.37\times$. Our experiments thus show that linear time, verification, and invertibility can be achieved without prohibitive cost.

\begin{credits}
\subsubsection{\ackname}
This study was funded in part by Swiss National Science Foundation grant number 200021\_219474 and State Secretariat for Education, Research, and Innovation / European Space Agency grant ``Embedded Flight Software Verification``.

We thank Romain Edelmann for useful discussions; Chapter 2 of his dissertation \cite{edelmann_efficient_2021} on lexing with zippers and derivatives provided a starting point for our zipper-based representation.

\subsubsection{\discintname}
The authors have no competing interests to declare that are relevant to the content of this article.

\end{credits}
\section{Data-Availability Statement}

This paper is accompanied by a Scala project including the full implementation of the lexer, the regular expression matching engine, and the memoization framework—including the mutable hash map implementation (and a copy of the LongMap implementation~\cite{chassot2024verifying}) together with all $k$-inductive specifications and lemmas. The project includes different example lexers for different grammars including a subset of Python, Amy (a teaching language), as well as the adversarial grammar described in Section~\ref{sec:performance}. It also contains the JSON manipulation application example from Section~\ref{sec:usage-examples}. The entire project verifies using Stainless~\cite{epfl_larastainless_2026}. The project finally contains the benchmarking code to reproduce our performance evaluation, as well as scripts to run the evaluation and analyze the resulting data. 
The paper is also accompanied by 1.35GB of SMT-LIB files generated by Stainless during verification, which can be used as a source of benchmarks for expressive categories of SMT competitions. The artifact consists of a Docker image containing all this material and all required software and is available on Zenodo~\cite{chassot_formally_2026}. The version number 4 on Zenodo (\url{https://doi.org/10.5281/zenodo.20084365}) has been evaluated by the artifact evaluation committee. The Scala project is also available in the Bolts repository~\cite{epfl_larabolts_2026} under \lstinline{lexers/regex}.
Some of the improved data structures such as the immutable arrays with a zero-copy slice operation, and the executable specification of the hash table, \lstinline{ListMap}, are now part of the Stainless standard library~\cite{epfl_larastainless_2026} under \lstinline{frontends/library/stainless}.

\bibliographystyle{splncs04}
\bibliography{references}

@String{Computing = "Computing" }

@String{Computer = "{IEEE} Computer" }

@String{Springer = "Springer-Verlag" }

@article{brzozowski1964derivatives,
  author       = {Janusz A. Brzozowski},
  title        = {Derivatives of Regular Expressions},
  journal      = {J. {ACM}},
  volume       = {11},
  number       = {4},
  pages        = {481--494},
  year         = {1964},
  url          = {https://doi.org/10.1145/321239.321249},
  doi          = {10.1145/321239.321249},
  bibsource    = {dblp computer science bibliography, https://dblp.org}
}

@phdthesis{edelmann_efficient_2021,
	title = {Efficient {Parsing} with {Derivatives} and {Zippers}},
	url = {https://infoscience.epfl.ch/handle/20.500.14299/179767},
	language = {en},
	urldate = {2024-08-08},
	school = {EPFL},
	author = {Edelmann, Romain},
	year = {2021},
	doi = {10.5075/epfl-thesis-7357},
}

@article{huet1997zipper,
  title={The zipper},
  author={Huet, G{\'e}rard},
  journal={Journal of functional programming},
  volume={7},
  number={5},
  pages={549--554},
  year={1997},
  publisher={Cambridge University Press}
}

@techreport{cattell1978formalization,
  title={Formalization and automatic derivation of code generators.},
  author={Cattell, Roderic Geoffrey Galton},
  year={1978},
  journal = {},
  institution = {Carnegie-Mellon University, Dept.~of Computer Science}
}

@inproceedings{nipkow1998verified, address={Berlin, Heidelberg}, title={Verified lexical analysis}, ISBN={978-3-540-49801-8}, DOI={10.1007/BFb0055126}, abstractNote={This paper presents the development and verification of a (very simple) lexical analyzer generator that takes a regular expression and yields a functional lexical analyzer. The emphasis is on simplicity and executability. The work was carried out with the help of the theorem prover Isabelle/HOL.}, booktitle={Theorem Proving in Higher Order Logics}, publisher={Springer}, author={Nipkow, Tobias}, editor={Grundy, Jim and Newey, Malcolm}, year={1998}, pages={1–15}, language={en} }

@article{DBLP:journals/toplas/Reps98,
  author       = {Thomas W. Reps},
  title        = {"Maximal-munch" Tokenization in Linear Time},
  journal      = {{ACM} Trans. Program. Lang. Syst.},
  volume       = {20},
  number       = {2},
  pages        = {259--273},
  year         = {1998},
  doi          = {10.1145/276393.276394},
}

@article{ouedraogo2023coqlex,
  title={Coqlex: Generating Formally Verified Lexers},
  author={Ouedraogo, Wendlasida and Scherer, Gabriel and Strassburger, Lutz},
  journal={The Art, Science, and Engineering of Programming},
  volume={8},
  number={1},
  pages={3--1},
  year={2023},
  publisher={AOSA, Inc.}
}

@inproceedings{egolf2022verbatim++, 
    address={Philadelphia PA USA}, 
    title={Verbatim++: verified, optimized, and semantically rich lexing with derivatives}, 
    ISBN={978-1-4503-9182-5}, url={https://dl.acm.org/doi/10.1145/3497775.3503694}, 
    DOI={10.1145/3497775.3503694}, 
    abstractNote={Lexers and parsers are attractive targets for attackers because they often sit at the boundary between a software system's internals and the outside world. Formally verified lexers can reduce the attack surface of these systems, thus making them more secure.}, 
    booktitle={Proceedings of the 11th ACM SIGPLAN International Conference on Certified Programs and Proofs}, 
    publisher={ACM}, 
    author={Egolf, Derek and Lasser, Sam and Fisher, Kathleen}, 
    year={2022}, 
    month=jan, 
    pages={27–39}, 
    language={en} 
}

@inproceedings{chattopadhyay_verified_2025,
	address = {New York, NY, USA},
	series = {{CPP} '25},
	title = {Verified and {Efficient} {Matching} of {Regular} {Expressions} with {Lookaround}},
	isbn = {979-8-4007-1347-7},
	url = {https://dl.acm.org/doi/10.1145/3703595.3705884},
	doi = {10.1145/3703595.3705884},
	urldate = {2025-01-24},
	booktitle = {Proceedings of the 14th {ACM} {SIGPLAN} {International} {Conference} on {Certified} {Programs} and {Proofs}},
	publisher = {Association for Computing Machinery},
	author = {Chattopadhyay, Agnishom and Li, Angela W. and Mamouras, Konstantinos},
	month = jan,
	year = {2025},
	pages = {198--213},
}

@article{varatalu2025re,
  title={RE\#: High Performance Derivative-Based Regex Matching with Intersection, Complement, and Restricted Lookarounds},
  author={Varatalu, Ian Erik and Veanes, Margus and Ernits, Juhan},
  journal={Proceedings of the ACM on Programming Languages},
  volume={9},
  number={POPL},
  pages={1--32},
  year={2025},
  publisher={ACM New York, NY, USA}
}

@article{owens2009regular,
  title={Regular-expression derivatives re-examined},
  author={Owens, Scott and Reppy, John and Turon, Aaron},
  journal={Journal of Functional Programming},
  volume={19},
  number={2},
  pages={173--190},
  year={2009},
  publisher={Cambridge University Press}
}

@inproceedings{chassot2024verifying, address={Cham}, title={Verifying a Realistic Mutable Hash Table}, ISBN={978-3-031-63498-7}, DOI={10.1007/978-3-031-63498-7_18}, abstractNote={In this work, we verify, using the Stainless program verifier, the mutable LongMap from the Scala standard library, a hash table using open addressing within a single array. As an executable specification, we write an immutable map based on a list of tuples and verify it against the mathematical definition of a map. We then show that LongMap's operations correspond to operations of this association list. To express the resizing of the hash table array, we introduce a new reference-swapping construct in Stainless. This allows us to apply the decorator design pattern without introducing aliasing. Our verification effort led us to find and fix a bug in the original implementation that manifests for large hash tables. Our performance analysis shows the verified version to be within a 1.5 factor of the original data structure.}, booktitle={Automated Reasoning}, publisher={Springer Nature Switzerland}, author={Chassot, Samuel and Kunčak, Viktor}, editor={Benzmüller, Christoph and Heule, Marijn J.H. and Schmidt, Renate A.}, year={2024}, pages={304–314}, language={en} }

@misc{coqlex_comparisonjson_2023_repo,
	title = {Coqlex Artifact Gitlab Repository},
	howpublished = {\url{https://gitlab.inria.fr/wouedrao/coqlex/-/tree/master/Comparison/JSON?ref_type=heads}},
	language = {en},
	urldate = {2025-09-25},
	journal = {GitLab},
    author={Ouedraogo, Wendlasida and Scherer, Gabriel and Strassburger, Lutz},
	month = dec,
	year = {2023},
}

@inproceedings{egolf_verbatim_2021,
	address = {San Francisco, CA, USA},
	title = {Verbatim: {A} {Verified} {Lexer} {Generator}},
	isbn = {978-1-6654-3732-5},
	shorttitle = {Verbatim},
	url = {https://ieeexplore.ieee.org/document/9474322/},
	doi = {10.1109/SPW53761.2021.00022},
	language = {en},
	urldate = {2023-05-11},
	booktitle = {2021 {IEEE} {Security} and {Privacy} {Workshops} ({SPW})},
	publisher = {IEEE},
	author = {Egolf, Derek and Lasser, Sam and Fisher, Kathleen},
	month = may,
	year = {2021},
	pages = {92--100},
}

@phdthesis{leroy_compcert_2024,
	type = {report},
	title = {The {CompCert} {C} verified compiler: {Documentation} and user's manual},
	shorttitle = {The {CompCert} {C} verified compiler},
	url = {https://inria.hal.science/hal-01091802},
	language = {en},
	urldate = {2025-09-30},
	school = {Inria},
	author = {Leroy, Xavier},
	month = dec,
	year = {2024},
	note = {Pages: 1},
}

@article{MadhavanKuncak17Memoization, title={Contract-based resource verification for higher-order functions with memoization}, volume={52}, ISSN={0362-1340}, DOI={10.1145/3093333.3009874}, abstractNote={We present a new approach for specifying and verifying resource utilization of higher-order functional programs that use lazy evaluation and memoization. In our approach, users can specify the desired resource bound as templates with numerical holes e.g. as steps ≤ ? * size(l) + ? in the contracts of functions. They can also express invariants necessary for establishing the bounds that may depend on the state of memoization. Our approach operates in two phases: first generating an instrumented first-order program that accurately models the higher-order control flow and the effects of memoization on resources using sets, algebraic datatypes and mutual recursion, and then verifying the contracts of the first-order program by producing verification conditions of the form ∃ ∀ using an extended assume/guarantee reasoning. We use our approach to verify precise bounds on resources such as evaluation steps and number of heap-allocated objects on 17 challenging data structures and algorithms. Our benchmarks, comprising of 5K lines of functional Scala code, include lazy mergesort, Okasaki's real-time queue and deque data structures that rely on aliasing of references to first-class functions; lazy data structures based on numerical representations such as the conqueue data structure of Scala's data-parallel library, cyclic streams, as well as dynamic programming algorithms such as knapsack and Viterbi. Our evaluations show that when averaged over all benchmarks the actual runtime resource consumption is 80% of the value inferred by our tool when estimating the number of evaluation steps, and is 88% for the number of heap-allocated objects.}, number={1}, journal={SIGPLAN Not.}, author={Madhavan, Ravichandhran and Kulal, Sumith and Kuncak, Viktor}, year={2017}, month=jan, pages={330–343} }

@inproceedings{DBLP:conf/oopsla/BrachthauserRO16,
  author       = {Jonathan Immanuel Brachth{\"{a}}user and
                  Tillmann Rendel and
                  Klaus Ostermann},
  editor       = {Eelco Visser and
                  Yannis Smaragdakis},
  title        = {Parsing with first-class derivatives},
  booktitle    = {Proceedings of the 2016 {ACM} {SIGPLAN} International Conference on
                  Object-Oriented Programming, Systems, Languages, and Applications,
                  {OOPSLA} 2016, part of {SPLASH} 2016, Amsterdam, The Netherlands,
                  October 30 - November 4, 2016},
  pages        = {588--606},
  publisher    = {{ACM}},
  year         = {2016},
  url          = {https://doi.org/10.1145/2983990.2984026},
  doi          = {10.1145/2983990.2984026},
  bibsource    = {dblp computer science bibliography, https://dblp.org}
}

@article{Rendel_2010_parsing_pretty_printing,
author = {Rendel, Tillmann and Ostermann, Klaus},
title = {Invertible syntax descriptions: unifying parsing and pretty printing},
year = {2010},
issue_date = {November 2010},
publisher = {Association for Computing Machinery},
address = {New York, NY, USA},
volume = {45},
number = {11},
issn = {0362-1340},
url = {https://doi.org/10.1145/2088456.1863525},
doi = {10.1145/2088456.1863525},
journal = {SIGPLAN Not.},
month = sep,
pages = {1–12},
numpages = {12}
}

@inproceedings{bucev2025formally,
  author       = {Mario Bucev and
                  Samuel Chassot and
                  Simon Felix and
                  Filip Schramka and
                  Viktor Kuncak},
  editor       = {Shankaranarayanan Krishna and
                  Sriram Sankaranarayanan and
                  Ashutosh Trivedi},
  title        = {Formally Verifiable Generated {ASN.1/ACN} Encoders and Decoders: {A}
                  Case Study},
  booktitle    = {Verification, Model Checking, and Abstract Interpretation - 26th International
                  Conference, {VMCAI} 2025, Denver, CO, USA, January 20-21, 2025, Proceedings,
                  Part {II}},
  series       = {Lecture Notes in Computer Science},
  pages        = {185--207},
  publisher    = {Springer},
  year         = {2025},
  url          = {https://doi.org/10.1007/978-3-031-82703-7\_9},
  doi          = {10.1007/978-3-031-82703-7\_9},
  bibsource    = {dblp computer science bibliography, https://dblp.org}
}

@inproceedings{leroy:hal-01238879,
  TITLE = {{CompCert - A Formally Verified Optimizing Compiler}},
  AUTHOR = {Leroy, Xavier and Blazy, Sandrine and K{\"a}stner, Daniel and Schommer, Bernhard and Pister, Markus and Ferdinand, Christian},
  URL = {https://inria.hal.science/hal-01238879},
  BOOKTITLE = {{ERTS 2016: Embedded Real Time Software and Systems, 8th European Congress}},
  ADDRESS = {Toulouse, France},
  ORGANIZATION = {{SEE}},
  YEAR = {2016},
  MONTH = Jan,
  HAL_ID = {hal-01238879},
  HAL_VERSION = {v1},
  publisher = {HAL},
  pages = {0, 0},
  comment = { pages are here just to suppress warning }
}

@inproceedings{edelmann_2020PLDI_zippy,
  author       = {Romain Edelmann and
                  Jad Hamza and
                  Viktor Kuncak},
  editor       = {Alastair F. Donaldson and
                  Emina Torlak},
  title        = {Zippy {LL(1)} parsing with derivatives},
  booktitle    = {Proceedings of the 41st {ACM} {SIGPLAN} International Conference on
                  Programming Language Design and Implementation, {PLDI} 2020, London,
                  UK, June 15-20, 2020},
  pages        = {1036--1051},
  publisher    = {{ACM}},
  year         = {2020},
  url          = {https://doi.org/10.1145/3385412.3385992},
  doi          = {10.1145/3385412.3385992},
  bibsource    = {dblp computer science bibliography, https://dblp.org}
}

@inproceedings{DBLP:conf/cav/GuilloudP25,
  author       = {Simon Guilloud and
                  Cl{\'{e}}ment Pit{-}Claudel},
  editor       = {Ruzica Piskac and
                  Zvonimir Rakamaric},
  title        = {Verified and Optimized Implementation of Orthologic Proof Search},
  booktitle    = {Computer Aided Verification - 37th International Conference, {CAV}
                  2025, July 23-25, 2025, Proceedings, Part {III}},
  address = {Zagreb, Croatia},
  series       = {Lecture Notes in Computer Science},
  volume       = {15933},
  pages        = {130--152},
  publisher    = {Springer},
  year         = {2025},
  url          = {https://doi.org/10.1007/978-3-031-98682-6\_8},
  doi          = {10.1007/978-3-031-98682-6\_8},
}

@article{DBLP:journals/pacmpl/DarraghA20,
  author       = {Pierce Darragh and
                  Michael D. Adams},
  title        = {Parsing with zippers (functional pearl)},
  journal      = {Proc. {ACM} Program. Lang.},
  volume       = {4},
  number       = {{ICFP}},
  pages        = {108:1--108:28},
  year         = {2020},
  url          = {https://doi.org/10.1145/3408990},
  doi          = {10.1145/3408990},
  bibsource    = {dblp computer science bibliography, https://dblp.org}
}

@article{DBLP:journals/pacmpl/SantoBP24,
  author       = {No{\'{e}} De Santo and
                  Aur{\`{e}}le Barri{\`{e}}re and
                  Cl{\'{e}}ment Pit{-}Claudel},
  title        = {A {Coq} Mechanization of {JavaScript} Regular Expression Semantics},
  journal      = {Proc. {ACM} Program. Lang.},
  volume       = {8},
  number       = {{ICFP}},
  pages        = {1003--1031},
  year         = {2024},
  url          = {https://doi.org/10.1145/3674666},
  doi          = {10.1145/3674666},
  bibsource    = {dblp computer science bibliography, https://dblp.org}
}

@Book{ranta-2011,
  author       = {Aarne Ranta},
  title        = {Grammatical Framework - Programming with Multilingual Grammars},
  series       = {{CSLI} Studies in Computational Linguistics},
  publisher    = {Cambridge University Press},
  year         = {2011},
  url          = {http://cslipublications.stanford.edu/site/9781575866277.shtml},
  isbn         = {978-1-57586-626-0},
  bibsource    = {dblp computer science bibliography, https://dblp.org}
}

@article{paxson1995flex,
  title={Flex--fast lexical analyzer generator},
  author={Paxson, Vern and others},
  journal={Lawrence Berkeley Laboratory},
  year={1995}
}

@article{Li_Mamouras_2025, 
  author       = {Angela W. Li and
                  Konstantinos Mamouras},
  title        = {Efficient Algorithms for the Uniform Tokenization Problem},
  journal      = {Proc. {ACM} Program. Lang.},
  volume       = {9},
  number       = {{OOPSLA1}},
  pages        = {1492--1518},
  year         = {2025},
  url          = {https://doi.org/10.1145/3720498},
  doi          = {10.1145/3720498},
  bibsource    = {dblp computer science bibliography, https://dblp.org}
}

@inproceedings{DBLP:conf/fmcad/BucevK22,
  author       = {Mario Bucev and
                  Viktor Kun\v{c}ak},
  editor       = {Alberto Griggio and
                  Neha Rungta},
  title        = {Formally Verified Quite {OK} Image Format},
  booktitle    = {22nd Formal Methods in Computer-Aided Design, {FMCAD} 2022, Trento,
                  Italy, October 17-21, 2022},
  pages        = {343--348},
  publisher    = {{IEEE}},
  year         = {2022},
  doi          = {10.34727/2022/ISBN.978-3-85448-053-2\_41},
}

@misc{jmhWebSite,
  key = {JMH},
  title = {Java Microbenchmark Harness ({JMH})},
  note = {Retrieved 25 January 2026},
  howpublished = {\url{https://github.com/openjdk/jmh}}
}

@misc{Ocamllex,
  key = {Ocamllex},
  title = {{OCamllex}},
  note = {Retrieved 15 May 2026},
  howpublished = {\url{https://ocaml.org/manual/5.3/lexyacc.html}}
}

@article{DBLP:journals/pacmpl/HamzaVK19,
  author       = {Jad Hamza and
                  Nicolas Voirol and
                  Viktor Kuncak},
  title        = {System {FR:} formalized foundations for the {S}tainless verifier},
  journal      = {Proc. {ACM} Program. Lang.},
  volume       = {3},
  number       = {{OOPSLA}},
  pages        = {166:1--166:30},
  year         = {2019},
  url          = {https://doi.org/10.1145/3360592},
  doi          = {10.1145/3360592},
  bibsource    = {dblp computer science bibliography, https://dblp.org}
}

@inproceedings{DBLP:conf/pldi/WurthingerWHWSS17,
  author       = {Thomas W{\"{u}}rthinger and
                  Christian Wimmer and
                  Christian Humer and
                  Andreas W{\"{o}}{\ss} and
                  Lukas Stadler and
                  Chris Seaton and
                  Gilles Duboscq and
                  Doug Simon and
                  Matthias Grimmer},
  editor       = {Albert Cohen and
                  Martin T. Vechev},
  title        = {Practical partial evaluation for high-performance dynamic language
                  runtimes},
  booktitle    = {Proceedings of the 38th {ACM} {SIGPLAN} Conference on Programming
                  Language Design and Implementation, {PLDI} 2017, Barcelona, Spain,
                  June 18-23, 2017},
  pages        = {662--676},
  publisher    = {{ACM}},
  year         = {2017},
  url          = {https://doi.org/10.1145/3062341.3062381},
  doi          = {10.1145/3062341.3062381},
  bibsource    = {dblp computer science bibliography, https://dblp.org}
}

@misc{chassot_formally_2026,
	title = {Formally {Verified} {Linear}-{Time} {Invertible} {Lexing} ({Artifact})},
	url = {https://zenodo.org/records/19830775},
	doi = {10.5281/zenodo.19830775},
    howpublished = {\url{https://zenodo.org/records/19830775}},
	urldate = {2026-05-15},
	publisher = {Zenodo},
	author = {Chassot, Samuel and Kunčak, Viktor},
	month = may,
	year = {2026},
}

@misc{epfl_larabolts_2026,
    key = {Bolts},
	title = {epfl-lara/bolts},
	copyright = {Apache-2.0},
	url = {https://github.com/epfl-lara/bolts},
    howpublished = {\url{https://github.com/epfl-lara/bolts}},
	urldate = {2026-05-15},
	publisher = {EPFL-LARA},
	month = apr,
	year = {2026},
	note = {original-date: 2017-10-12T10:48:10Z},
}

@misc{epfl_larastainless_2026,
    key = {Stainless},
	title = {epfl-lara/stainless},
	copyright = {Apache-2.0},
	url = {https://github.com/epfl-lara/stainless},
    howpublished = {\url{https://github.com/epfl-lara/stainless}},
	urldate = {2026-05-15},
	publisher = {EPFL-LARA},
	month = may,
	year = {2026},
	note = {original-date: 2016-08-17T14:08:38Z},
}

\newpage
\appendix

\section{Regular Expression Engine: Additional Details}
\label{app:regexp-engine}

This appendix complements Section~\ref{sec:regexp-engine} with additional definitions and proof-oriented details.

\subsection{Definition and Implementation}
\label{app:regexp-definition-implementation}

As mentioned in Section~\ref{sec:usage-examples}, the alphabet is arbitrary and defined by a type parameter. We call elements of the alphabet \emph{characters} and call \emph{strings} ordered sequences of characters.

Regular expressions are defined by the usual constructs: \textit{EmptyExpr} for $\epsilon$, \textit{EmptyLang} for $\emptyset$, \textit{ElementMatch(a)}, \textit{Union}, \textit{Concat}, and \textit{Star}. We use $\lambda$ to denote the empty string.

The function $\matchR(r,s)$ decides membership in the language of $r$ using Brzozowski derivatives~\cite{brzozowski1964derivatives}. 
As shown by Brzozowski~\cite{brzozowski1964derivatives}, matching a string and a regular expression is equivalent to checking for nullability (i.e., $\lambda \in L$) of its derivative: $z \in L \Leftrightarrow \lambda \in \partial_z L$, where $L$ is the language represented by the regular expression. The derivative of a language $L$ with respect to a character $a$, is a new language composed of all suffixes of words of $L$ that start with $a$. Figure~\ref{fig:derivatives-theory-definitions} shows the formal definition of derivatives for languages, which can be extended to regular expressions.
The $\matchR$ function therefore computes the derivative with respect to each character and checks for nullability.

\begin{figure}
    \centering
    \begin{align*}
        \partial_a L &= \{z \mid a::z\in L\}\\
        \partial_\lambda L &= L\\
        \partial_{a::z} L &= \partial_z(\partial_a L) 
    \end{align*}
    \caption{Definition of Brzozowski's derivatives where $a$ is a character, $z$ and $\lambda$ are strings, and $L$ is a language.} \Description{Formal definition of Brzozowski's derivatives where $a$ is a character, $z$ and $\lambda$ are strings, and $L$ is a language.}
    \label{fig:derivatives-theory-definitions}
\end{figure}

The engine also provides $\findLongestMatch$, which computes the longest prefix of a string matched by $r$. While it could be implemented by testing all prefixes with $\matchR$, our implementation traverses the input once and stops early when the current derived expression becomes equivalent to $\emptyset$, thus computing at most one derivative per input character. If no prefix matches, the empty string is returned. We do not need to handle nullable rules, as lexer rules are required to be non-nullable.

\smartparagraph{Specification}
We encode an executable specification $\matchRSpec$ for $\matchR$ (Appendix Section~\ref{subsec:matchRSpec}) and prove the equivalence:
\[
\forall r,s:\ \matchR(r,s) = \matchRSpec(r,s).
\]

% The main discussion of zippers is in Section~\ref{sec:regexp-engine}.
% We provide additional depth in Appendix~\ref{subsec:regex-zipper-optimization-depth}.
See Section~\ref{subsec:regex-zipper-optimization-depth} for additional detail on the zipper representation and its derivative operations.

Figure~\ref{fig:zipper-adt-def-focus-unfocus-signature} shows our zipper representation. Intuitively, a \lstinline{Context} represents a concatenation (e.g., $\Context(r_1,r_2,r_3,r_4)$ corresponds to $r_1\cdot(r_2\cdot(r_3\cdot r_4))$), and a zipper represents a union of such contexts. The matching function $\matchZ$ follows the same structure as $\matchR$: it derives with respect to each character and checks nullability.

\smartparagraph{Specification}
Zipper-based matching is designed as a drop-in replacement for regular expression matching and is specified by:
\[
\forall s,r,z:\ \unfocus(z)=r \implies \matchR(r,s)=\matchZ(z,s).
\]
We proved this equivalence in Stainless.

\subsection{Ziplex Design}
\label{subsec:ziplex-design-appendix}
Figure~\ref{fig:tokens-rules-adt-definition-lex-sig} shows the Algebraic Data Type (ADT) definition for tokens. In addition to its semantic value, a token also contains the \emph{rule} that produced it, a \emph{size} which corresponds to the length of the string originally consumed, and a ghost value of the original string (used in the proof but erased at run time). The token class maintains an invariant stating that applying the back transformation to the semantic value gives back the original string, and that the size is consistent. Finally, the token class provides a method computing this original string at runtime.
\begin{figure}
    \centering
    \begin{lstlisting}
case class IntegerValue(value: BigInt, text: List[Char]) extends TokenValue:
object IntegerValueUtils:
  @pure def charToBigInt(c: Char): BigInt = c match
    case _ if c == '1' => 1 // up to 9
    case _             => 0      
  @pure def charsToBigInt(l: List[Char], acc: BigInt = 0): BigInt = 
    if l.isEmpty then acc else charsToBigInt(tl, acc * 10 + charToBigInt(hd))
        }
case object IntegerValueInjection:
  def toValue(v: Sequence[Char]): TokenValue =
    val list = v.efficientList
    IntegerValue(IntegerValueUtils.charsToBigInt(list), list)
  def toCharacters(t: TokenValue): Sequence[Char] = t match
    case IntegerValue(_, text) => seqFromList(text)
    case _                     => emptySeq()
val injection: TokenValueInjection[Char] = TokenValueInjection(toValue, toCharacters)
    \end{lstlisting}
    \caption{Token semantic value for JSON integer literal along with the injective transformation function.}
    \Description{Token semantic value for JSON integer literal along with the injective transformation function.}
    \label{fig:json-lexer-integer-semantic-value}
\end{figure}

% removed from inject above:
% ghostExpr:
%                 assert:
%                   unfold(semiInverseBody(toCharacters, toValue))
%                   semiInverseBody(toCharacters, toValue)
%                 unfold(semiInverse(toCharacters, toValue))

\begin{figure}
    \centering
    \begin{lstlisting}
case class Rule[C](regex: Regex[C], tag: String, isSep: Boolean, trsfm: TokenValueInjection[C])
case class Token[C](value: TokenValue, rule: Rule[C], size: BigInt, @ghost orgnlChars: List[C]):
  require(!orgnlChars.isEmpty)
  require(orgnlChars == rule.trsfm.toChars(value).list)
  require(size == orgnlChars.size)
def lex[C: ClassTag](rules: List[Rule[C]], input: Sequence[C]): (Sequence[Token[C]], Sequence[C])
def print[C: ClassTag](v: Sequence[Token[C]]): Sequence[C]
    \end{lstlisting}
    \caption{ADT definitions for tokens and rules of the lexer framework. Signature of the \textit{lex} function.}
    \Description{ADT definitions for tokens and rules of the lexer framework}
    \label{fig:tokens-rules-adt-definition-lex-sig}
\end{figure}

\noindent\textbf{Rule definitions}: One lexer rule is composed of a regular expression, a user-defined name (a \emph{tag}), a flag indicating whether the rule is for a "separator" token or not (Section~\ref{subsec:invertibility-separator-tokens}), and a semantic value transformation defined at the previous step. Figure~\ref{fig:tokens-rules-adt-definition-lex-sig} shows ADT definitions as Scala case classes for rules. Section~\ref{sec:regexp-engine} details the regular expression implementation and verification.
At this step, users then define the regular expressions %(Section~\ref{sec:regexp-engine}) 
for the rules. The available constructs include the usual single element of the alphabet, concatenation, union, and Kleene star. Figure~\ref{fig:json-lexer-integer-rule} shows the regular expression and rule definition for integer-literal value tokens. Note that the regular expression definition uses user-defined syntactic sugar functions for readability.
\begin{figure}
    \centering
    \begin{lstlisting}
def intRe: Regex[Char] = opt('-'.r) ~ ('0'.r.+ | anyOf("123456789") ~ anyOf("0123456789").*)
val integerLiteralRule = Rule(r=intRe, tag="integerLiteral", isSep=false, 
trnsfm=IntegerValueInjection.injection)
    \end{lstlisting}
    \caption{Rule definition for integer literal in the JSON lexer. The regular expression is expressed using user-defined syntactic sugars.}
    \Description{Rule definition for integer literal in the JSON lexer. The regular expression is expressed using user-defined syntactic sugars.}
    \label{fig:json-lexer-integer-rule}
\end{figure}

\noindent\textbf{Rule priorities}: Users group rules in a list whose ordering defines the rule priorities (closer to the head of the list means higher priority).

\noindent\textbf{Rule validation and lexing}: at this point the lexer is ready to lex. For Stainless to accept the code, users need to ensure that no regular expression is nullable and that the tags are unique.

Other printing functions are presented in Section~\ref{subsec:invertibility-separator-tokens}.

\subsection{Formal Specification of Lexer with Maximal Munch Policy}

The correctness specification for \ZipLex is the maximal munch principle, or longest match semantic, first mentioned in \cite{cattell1978formalization}. In lexical analysis, this principle states that each produced token should be the longest possible given the input string and the rules. We give the Stainless formalism of the longest match semantic in Figure~\ref{fig:max-munch-stainless-theorem}. This theorem states that, for a given tokenization, either no tokens are produced, or the first token $t = \tkn{z}$ is produced by a given rule $r$; in the latter case, no rules with a smaller index match another prefix $z'$ with $z.size \leq z'.size$, and no other rule matches a prefix $z''$ with $z.size < z''.size$. In other words, no rules match a longer prefix than the one produced, and no other rule with a smaller index matches the produced prefix. This theorem proves that the \textit{lex} function follows the maximal munch specification for the first token and, given the recursive nature of the lexer, proves that it follows this specification for all tokens. 
%  Move to a later section:
%The proof is done by induction with a set of intermediate lemmas about lists and the \lstinline{maxPrefix} and \lstinline{maxPrefixOneRule} functions.

\begin{figure}
    \centering  
    \begin{lstlisting}
def theoremLexSoundFirstChar[C](
    rules: List[Rule[C]], 
    input: List[C], suffix: List[C], 
    tokens: List[Token[C]], 
    r: Rule[C], otherR: Rule[C], otherP: List[C]): Unit = {
  require(rules.contains(r) && rules.contains(otherR))
  require(lex(rules, input) == (tokens, suffix))    
  require(tokens.isEmpty || tokens.head.characters.size <= otherP.size)
  require(tokens.isEmpty || tokens.head.tag == r.tag)
  require(tokens.isEmpty || tokens.head.isSeparator == r.isSeparator)
  require(ListUtils.isPrefix(otherP, input))
  require(r != otherR)
  // ...
} ensuring:
    if (ListUtils.getIndex(rules, otherR) < ListUtils.getIndex(rules, r)) 
    then !matchR(otherR.regex, otherP)
    else tokens.size > 0 && otherP.size <= tokens.head.characters.size || 
        !matchR(otherR.regex, otherP)
    \end{lstlisting}
    \caption{Correctness theorem (in Stainless) for the \textit{lex} function, expressing the longest match semantic for lexical analysis}
    \Description{Correctness theorem for the lex function, expressing the longest match semantic for lexical analysis, in the Stainless formalism.}
    \label{fig:max-munch-stainless-theorem}
\end{figure}

\subsection{Simple Approach: Separator Tokens}
\label{subsec:invertibility-separator-tokens}

For completeness, we also present a simple yet restrictive approach for ensuring separability. The approach introduces two different classes of tokens, \emph{separator tokens} and \emph{non-separator tokens}. We design those classes such that interleaving them in a sequence of tokens then avoids tokens collapsing. 
%To ensure this property, two tokens of different classes should never collapse together. 
This is achieved by restricting the characters each class can use in regular expressions. The set of characters used by a regular expression is defined as the set of characters that appear in an \lstinline{ElementMatch} node within the regular expression. The restriction is that the \emph{sets of used characters for the rules of each class must be disjoint}. In other words, characters used by separator tokens must not be used by non-separator tokens. In the case of a compiler's lexer, an example of separator tokens might be whitespace. In this case, no non-separator tokens can contain a whitespace character. We can then prove that a regular expression cannot match a string that contains one character which is not in its set.

If the rules follow this invariant, we can prove that two tokens of different classes printed next to each other will never produce a string that can be matched by any rule. More generally, we can prove that, given a sequence of tokens $ts = t_1, t_2,\dots,t_n$ and $s = s_1:::s_2:::\dots:::s_n$ the string obtained when printing it, if $t_1$ and $t_2$ each belong to a different class, the longest prefix of the string $s$ matched by any rule cannot be longer than $s_1$. Indeed, we can extract the string $s_1 ::: s_2(0)$ from $s$, and as $t_1$ and $t_2$ belong to different classes, their rules' regular expressions use disjoint sets of characters. The string $s_1 ::: s_2(0)$ contains characters belonging to separator and non-separator rules. Therefore, no rules can match a string starting with $s_1 ::: s_2(0)$, and thus the longest prefix matched by any rule can only be $s_1$.
Using these lemmas, we can prove that $\lex(\print(ts)) = ts$ if $ts$ is an interleaving of separator and non-separator tokens. 

In practice, maintaining this invariant might result in storing unnecessary separators, so we also propose another approach where separator tokens are ignored, that is, we compare token sequences modulo separator tokens. To simplify working such such token sequences, we implement a printing function that inserts a given separator token in between all tokens from the sequence to print, called $\printWithSep$. We prove $\lex(\printWithSep(ts)) = ts'$ where $ts$ and  $ts'$ are equal up to separator tokens. We also evaluated a version $\printWithSepWhenNeeded$ that inserts the separator token only if it is strictly needed. However, to determine if a separator token is needed, it tries to lex the string without inserting the separator token and compares it to the input, which is possibly less efficient than desired. 

Separator tokens are usable when the lexical syntax reserves a special character to be used for a separator, such a null character that cannot be present as part of tokens.
In a lexer for a typical programming language, however, it is common to include token classes such as string literals, which contain most characters of the input alphabet. This makes the above definition, which does not examine the order of characters within a token, inappropriate,  justifying a more flexible approach that we describe in the main body of the paper (Section~\ref{subsec:invertibility-r-path-predicate}). 

\section{Regular Expression Matching Engine Performance Evaluation}
\label{subsec:performance-regex-matching}

We evaluate the regular expression engine, both the naive and the zipper-based implementations, using \lstinline{List} and \lstinline{Sequence}, as well as with and without memoization.
To evaluate the matching engine, we use a regular expression from one of the example lexers we build for Amy, namely, the single-line comment regular expression. This regular expression matches all strings that start with "\texttt{//}" and do not contain newline characters, or formally $r = \texttt{/} \cdot \texttt{/} \cdot \Sigma_1^* $. 
%Figure \ref{fig:single-line-comment-regex} shows this regular expression in mathematical notation. 
We call the match function with this regular expression and randomly generated valid comment strings of varying length, from 5 to 120, in steps of 5. We compare the naive derivative-based matching, using \lstinline{List} and \lstinline{Sequence}, with and without memoization to zipper-based matching, also using \lstinline{List} and \lstinline{Sequence}, with and without memoization. 
%As a baseline, we use the Scala standard library implementation of regex\footnote{\url{https://www.scala-lang.org/api/2.13.4/scala/util/matching/Regex.html}}. 
Figure~\ref{fig:regex-performance-analysis} shows the results. First, we note that matching regular expressions with naive Brzozowski derivatives exhibits experimentally a quadratic complexity in the length of the input string (left plot), whereas the zipper-based implementation is empirically linear. We can see that using \lstinline{Sequence} or \lstinline{List} does not make a significant difference (right plot). Indeed, the matching algorithm traverses the string from left to right, which is also efficient using linked lists. The real benefit of \lstinline{Sequence} is seen when slicing and concatenating, which happens during the lexing process. On the other hand, memoization significantly increases performance (right plot), improving it by a factor of around 70. 
\begin{figure}
    \centering
    \begin{minipage}[t]{0.47\textwidth}
        \centering
        \includegraphics[width=\linewidth]{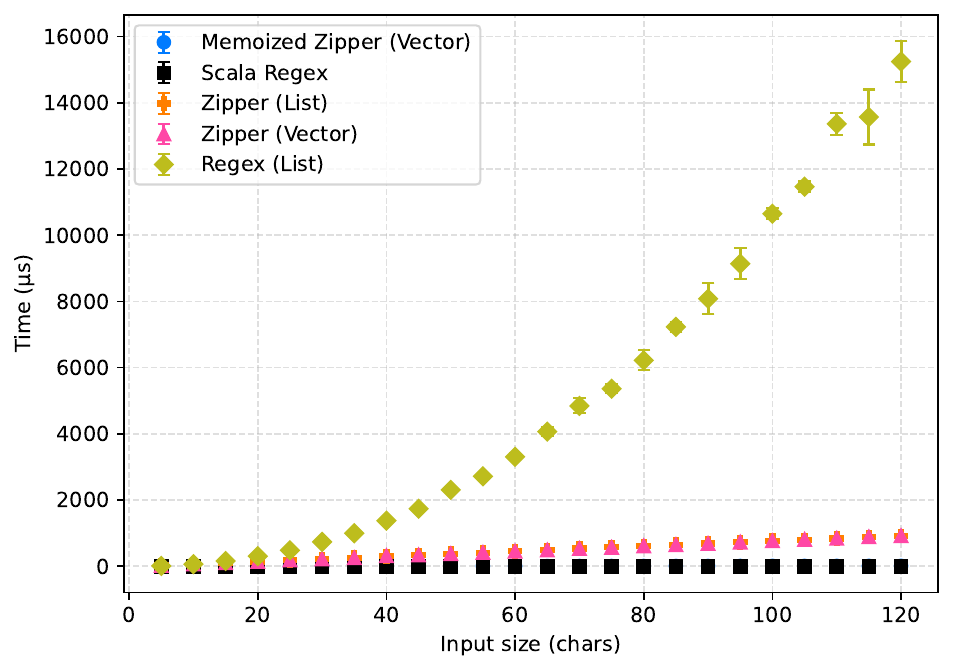}
    \end{minipage}
    \hfill
    \begin{minipage}[t]{0.46\textwidth}
        \centering
        \includegraphics[width=\linewidth]{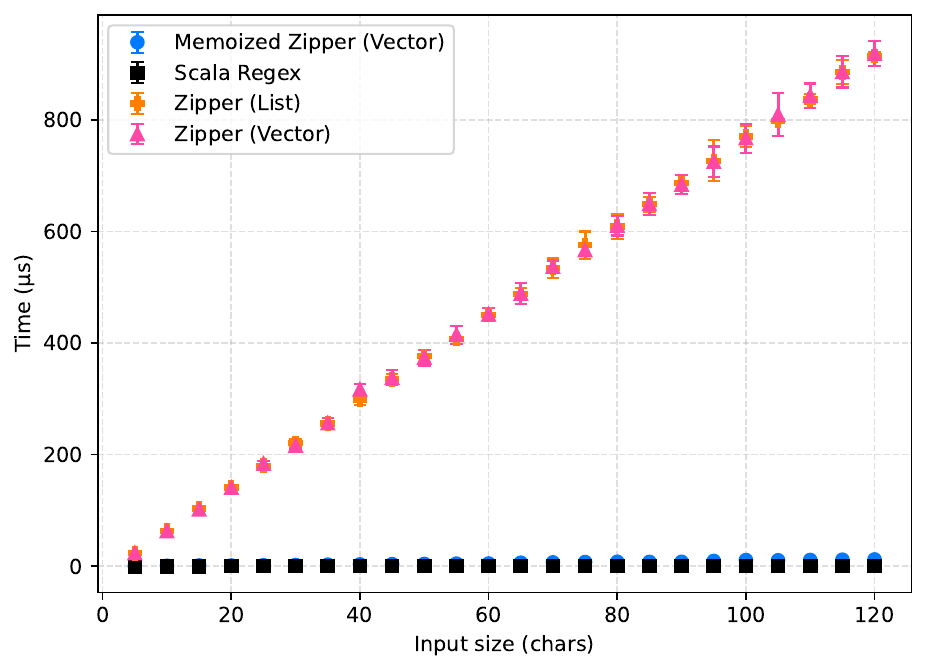}
    \end{minipage}

    % \vskip1em

    % \begin{minipage}[t]{0.49\textwidth}
    %     \centering
    %     \includegraphics[width=\linewidth]{res/Single_Line_Comments_Zipper_vs_Scala_Regex_Vector_with_Memoization_.pdf}
    % \end{minipage}
    % \hfill
    % \begin{minipage}[t]{0.49\textwidth}
    %     \centering
    %     \includegraphics[width=\linewidth]{res/Single_Line_Comments_Matching_Regex_List_with_Polynomial_Regression_deg_2_.pdf}
    % \end{minipage}
    \caption{Matching of single-line comments of size $N$. Both plots show the same operations, but some omit the slowest matching functions for better visibility.}
    \Description{Matching of single-line comments of size $N$. Both plots show the same operations, but some omit the slowest matching functions for better visibility.}
    \label{fig:regex-performance-analysis}
\end{figure}

\subsection{R-Path Based Predicate}

\subsubsection{Injection}
\label{app:injection}

Figure \ref{fig:stainless-library-injection} shows the definition of the \lstinline{Injection} construct added to Stainless library.

\begin{figure}
    \centering
    \begin{lstlisting}
        def semiInverse[X, Y](f: X => Y, g: Y => X): Boolean = 
          Forall((y: Y) => f(g(y)) == y)
        case class Injection[A, B](f: A => B, witness: B => A):
          require(semiInverse(witness, f)) // witness(f(a)) == a
          def apply(a: A): B = f(a)
          @ghost def lemmaInv(): Unit = {}.ensuring(_ => semiInverse(witness, f))
    \end{lstlisting}
    \caption{Definition of the new \lstinline{Injection} construct. In the token transformation case, \lstinline{f} is $\transform$ and \lstinline{witness} is $\charsOf$.}
    \Description{Definition of the new \lstinline{Injection} construct}
    \label{fig:stainless-library-injection}
\end{figure}

\subsubsection{PrintableTokens abstract type}
\label{app:printabletokens-interface}

Figure \ref{fig:printable-tokens-wrapper} shows the Scala definition of $\PrintableTokens$.

\begin{figure}[hbt]
    \centering
\begin{lstlisting}
case class PrintableTokens[C](rules: List[Rule[C]], tokens: Vector[Token[C]]):
  require(!rules.isEmpty)
  require(Lexer.rulesInvariant(rules))
  require(Lexer.rulesProduceEachTokenIndividually(rules, tokens))
  require(Lexer.separableTokens(tokens, rules))
  def size: BigInt = tokens.size
  def print(): Vector[C] = { // ...
  }.ensuring(res => Lexer.lex(rules, res) == (tokens, Vector.empty[C]))
  def append(other: PrintableTokens[C]): Option[PrintableTokens[C]] = {
    require(rules == other.rules)
    if (Lexer.separableTokensPredicate(tokens.last, other.tokens.head, rules)) then
      Some(PrintableTokens(rules, tokens ++ other.tokens))
    else None()
  }.ensuring(res => res.isEmpty || 
    (res.get.rules == rules && res.get.tokens == tokens ++ other.tokens && 
    res.get.print() == this.print() ++ other.print()))
  def slice(from: BigInt, to: BigInt): PrintableTokens[C] = {
    require(0 <= from && from <= to && to <= tokens.size) // ...
  }.ensuring(res => res.rules == rules && res.tokens == tokens.slice(from, to))
\end{lstlisting}
    \caption{$\PrintableTokens$ token sequence abstraction that maintains $\sep$ property on successive elements of the sequence. Also shown are operations {\tt append} and {\tt slice} that maintain the invariant.}
    \Description{Wrapper for printable token sequences (called $\PrintableTokens$) preserving the R-Path predicate $\sep(ts)$ invariant, along with operations preserving said invariant.}
    \label{fig:printable-tokens-wrapper}
\end{figure}

\subsection{$\matchR$ Specification}
\label{subsec:matchRSpec}
The specification of $\matchR$ is the following:
\begin{align*}
    match(\epsilon, s) &\iff s = \lambda \\
    match(\emptyset, s) &\iff false \\
    match( \llbracket a \rrbracket, s) &\iff s = "a"\\
    match(r_i^*, s) &\iff (s = \lambda)\ \vee \ \\
                    & \hspace{2em} (\exists (s_1, s_2)\ |\ s = s_1 \concat s_2\  \wedge\ match(r_i, s_1)\  \wedge\ match(r_i^* , s_2))\\
     match(r_1 + r_2, s) &\iff match(r_1, s)\ \vee\ match(r_2, s)\\
      match(r_1 \cdot r_2, s) &\iff \exists (s_1, s_2)\ |\ s = s_1 \concat s_2\  \wedge\ match(r_1, s_1)\  \wedge\ match(r_2 , s_2)
\end{align*}

\subsection{Zipper Optimization - Depth}
\label{subsec:regex-zipper-optimization-depth}

%Although elegant and simple, the implementation based on Brzozowski derivatives is inefficient. Indeed, the size of the regular expressions quickly grows during derivation if no simplification is performed. In addition, the recursive nature of the process leads to a high complexity in the size of the regular expression and redundant computations.
In this section, we explore the Zipper optimization described in Section \ref{subsec:regex-zipper-optimization}.

%For these reasons, we implement and verify an optimization based on Huet's zippers~\cite{huet1997zipper}. We follow the strategy proposed by Edelmann~\cite[Chapter 2]{edelmann_efficient_2021}, which provides a model and a proof in Rocq. We bridge the gap with implementation by providing a proof for a working Scala implementation, using Stainless. 
% The zipper-based representation has a comparative advantage over classical recursive regular expressions (Section~\ref{subsec:regexp-definition-implementation}) when it comes to memoization: the number of different zippers that can be visited during a derivation process is finite~\cite[Chapter 2]{edelmann_efficient_2021}. Classical recursive regular expressions indeed have a tendency to monotonically grow in size during derivation, leading to a large fraction of memoization cache misses. We return to this property when discussing memoization in Section~\ref{sec:memoization}.

In general, the zipper proposed by Huet~\cite{huet1997zipper} is a technique to traverse a tree-like data structure more efficiently. In particular, the main idea is to be able to move the "focus" to any node, as opposed to only the root node in a traditional tree-like data structure. For a binary tree, for example, a corresponding zipper is composed of a focus in the form of a binary tree (the subtree starting at the focused node) and a context that contains the information to reconstruct the original tree. More specifically, the context contains the nodes encountered when traversing from the root to the focused node and the neighbor subtrees that are not visited. Zippers offer an operation that reconstructs the original tree from a focused subtree and its context.

This pattern is interesting for the derivative operation that traverses potentially large tree-like structures of regular expressions. With zippers, instead of having to call constructors after recursive calls, these are pushed to contexts.
 
Figure~\ref{fig:zipper-adt-def-focus-unfocus-signature} shows the zipper representation we use in our regular expression engine.

In our zipper implementation, focus nodes are always implicit \lstinline{EmptyExpr}s. Also, the context is always the right neighbor and the parents of the current focus. We can restrict our zippers to this format because the focus moves only along the recursive structure of Brzozowski's derivatives. We do not further develop the justification for this structure in this work, as~\cite{edelmann_efficient_2021} explains it in depth. As explained in Section \ref{subsec:regex-zipper-optimization}, intuitively, contexts represent concatenations of their regular expressions: $\Context(r_1, r_2, r_3, r_4)$ is equivalent to $r_1 \cdot (r_2 \cdot (r_3 \cdot r_4))$. Zippers represent unions of contexts: $\Set(c_1, c_2, c_3)$ is equivalent to $r_{c_1} + (r_{c_2} + r_{c_3})$ where $r_{c_1}$, $r_{c_2}$, and $r_{c_3}$ are the regular expressions corresponding to each context.

The $\focus$ function creates a zipper from any regular expression, and its \emph{ghost} inverse $\unfocus$ reconstructs a regular expression from a zipper. Their signatures are shown in Figure~\ref{fig:zipper-adt-def-focus-unfocus-signature}. $\unfocus$ is \textit{ghost} because it has to work on a list of contexts, and not a set, to allow inductive reasoning about it. Note that $\focus$ ensures that the produced zipper can be unfocused to the same regular expression. This is not true in the other direction: the zipper $\Set(\Context(r_1, r_2), \Context(r_3, r_4))$ would be \textit{unfocused} to $(r_1\cdot r_2) + (r_3 \cdot r_4)$ and $\focus((r_1\cdot r_2) + (r_3 \cdot r_4)) = \Set(\Context((r_1\cdot r_2) + (r_3 \cdot r_4)))$. One should then note that $\unfocus$ is not injective, meaning that there exist multiple zippers that are \textit{unfocused} to the same regular expression. On the other hand, $\focus$ is injective.

\begin{figure}
    \centering
    \begin{lstlisting}[language=scala]
        def derivStepZUp[C](context: Context[C], a: C): Zipper[C] = // ...
        def derivStepZDown[C](expr: Regex[C], context: Context[C], a: C): Zipper[C] = // ...
        def derivStepZ[C](z: Zipper[C], a: C): Zipper[C] = z.flatMap(c => derivStepZUp(c, a))
        def matchZ[C](z: Zipper[C], input: List[C]): Boolean = 
          if (input.isEmpty) nullableZipper(z) else matchZ(derivStepZ(z, input.head), input.tail)
    \end{lstlisting}
    \caption{Brzozowski derivatives on zippers, and matching function for zippers.}
    \Description{Brzozowski derivatives on zippers, and matching function for zippers.}
    \label{fig:zippers-derivative-implementation}
\end{figure}

The derivation operation on zippers is implemented in two separate phases. \lstinline{derivationUp} takes a context and moves the focus up as long as the focused expression is nullable, i.e., visits all regular expressions that can be reached without consuming a character from the input. The \lstinline{derivationDown} phase moves the focus down the regular expressions, following a structure similar to the Brzozowski derivative. When reaching a leaf, it returns either a new zipper with the context in case the leaf is the right character, or an empty set. The empty set is used to represent the zipper matching the empty language. The derivative function signatures are shown in Figure~\ref{fig:zippers-derivative-implementation}.
The matching function for zippers $\matchZ$ follows the same algorithm defined for regular expression in Section~\ref{subsec:regexp-definition-implementation} and is also shown in Figure~\ref{fig:zippers-derivative-implementation}.

\subsection{Performance: Additional Results}

Figure \ref{fig:lexer-json-manipulation-lexing-and-R-path-with-without-mem} shows additional performance analysis results.

\begin{figure}[bth]
    \centering
    \begin{minipage}[t]{0.49\textwidth}
        \centering
        \includegraphics[width=\linewidth]{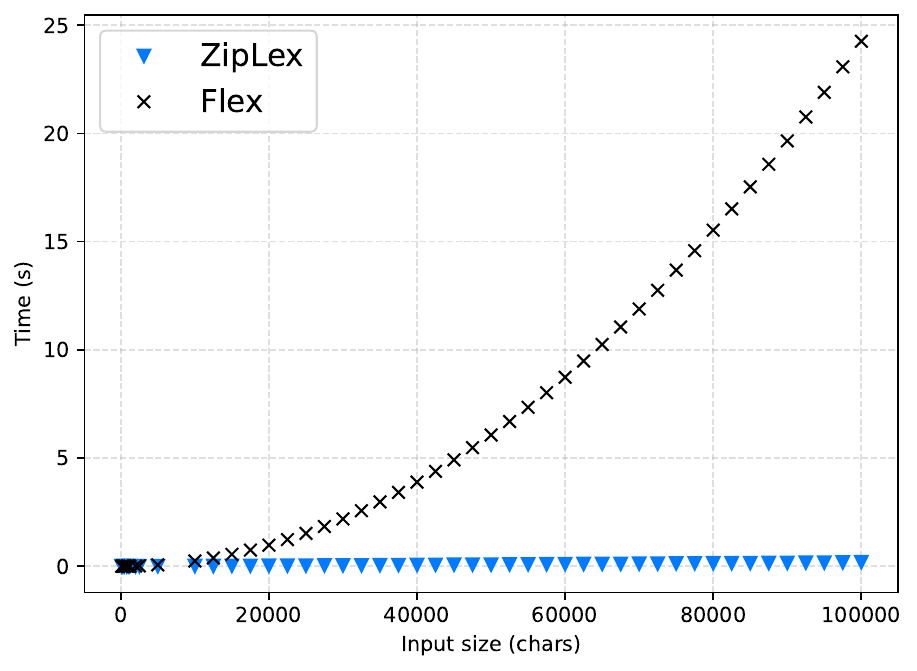}
    \end{minipage}

    \caption{
    Lexing time with grammar rules $r_1=a$ and $r_2=a^*b$, on string of $a$s of sizes ranging from 100 to 50K (or 100K), for lexers Ziplex, and flex.} 
    \label{fig:lexer-json-manipulation-lexing-and-R-path-with-without-mem}
\end{figure}

\subsection{Verification Effort: Verification Performance}

Figure \ref{fig:lexer-json-manipulation-lexing-and-R-path-with-without-mem} shows the distribution of VCs by solving time.

\begin{figure}[bth]
    \centering
    \begin{minipage}[t]{0.5\textwidth}
        \centering
        \includegraphics[width=\linewidth]{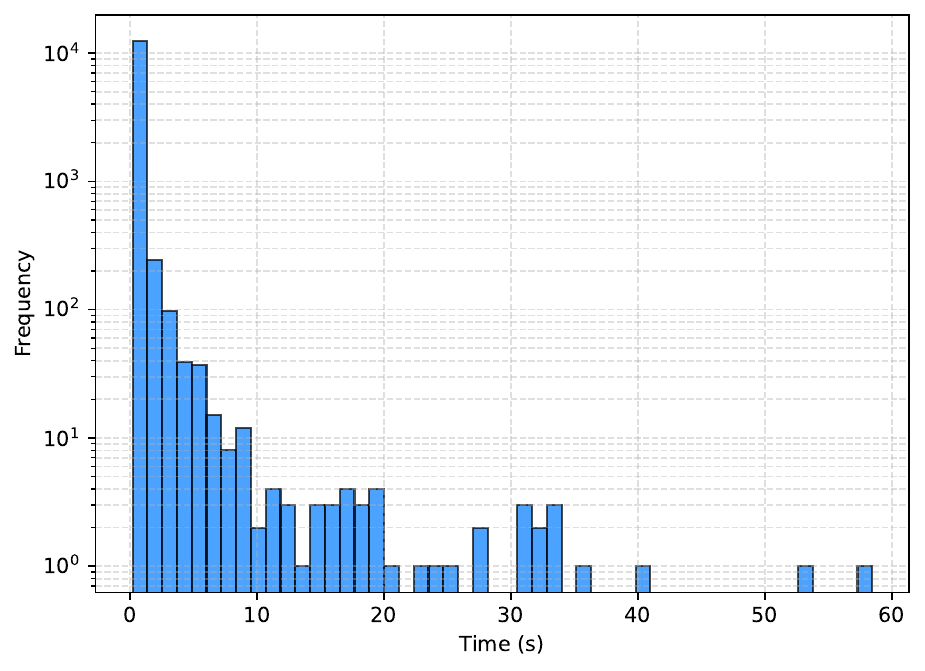}
    \end{minipage}
    \caption{Distribution of verification time for verification conditions with 100 buckets; the y-axis uses a logarithmic scale; } 
    \label{fig:vcs_time_distribution}
\end{figure}
\end{document}